\documentclass[journal=mamobx,manuscript=article]{achemso}
\setkeys{acs}{articletitle=true}

\SectionNumbersOn

\usepackage[version=3]{mhchem} % Formula subscripts using \ce{}
\usepackage[T1]{fontenc} % Use modern font encodings

\usepackage{color}
\usepackage{gensymb}
\usepackage {indentfirst}
\author{Xiaolei Xu}
\affiliation{State Key Laboratory of Polymer Physics and Chemistry, Changchun Institute of Applied Chemistry, Chinese Academy of Sciences, Changchun 130022, P. R. China}

\author{Jack F. Douglas}
\email{jack.douglas@nist.gov}
\affiliation{Materials Science and Engineering Division, National Institute of Standards and Technology, Gaithersburg, Maryland 20899, United States}

\author{Wen-Sheng Xu}
\email{wsxu@ciac.ac.cn}
\affiliation{State Key Laboratory of Polymer Physics and Chemistry, Changchun Institute of Applied Chemistry, Chinese Academy of Sciences, Changchun 130022, P. R. China}

\title{Generalized Entropy Theory Investigation of the Relatively High Segmental Fragility of Many Glass-Forming Polymers}

\abbreviations{IR,NMR,UV}

\begin{document}
	
%%%%%%%%%%%%%%%%%%%%%%%%%%%%%%%%%%%%%%%%%%%%%%%%%%%%%%%%%%%%%%%%%%%%%
%% The "tocentry" environment can be used to create an entry for the	%% graphical table of contents. It is given here as some journals	%% require that it is printed as part of the abstract page. It will
%% be automatically moved as appropriate.	%%%%%%%%%%%%%%%%%%%%%%%%%%%%%%%%%%%%%%%%%%%%%%%%%%%%%%%%%%%%%%%%%%%%%
\begin{tocentry}
		
 \centering
 \includegraphics[height=4.5cm]{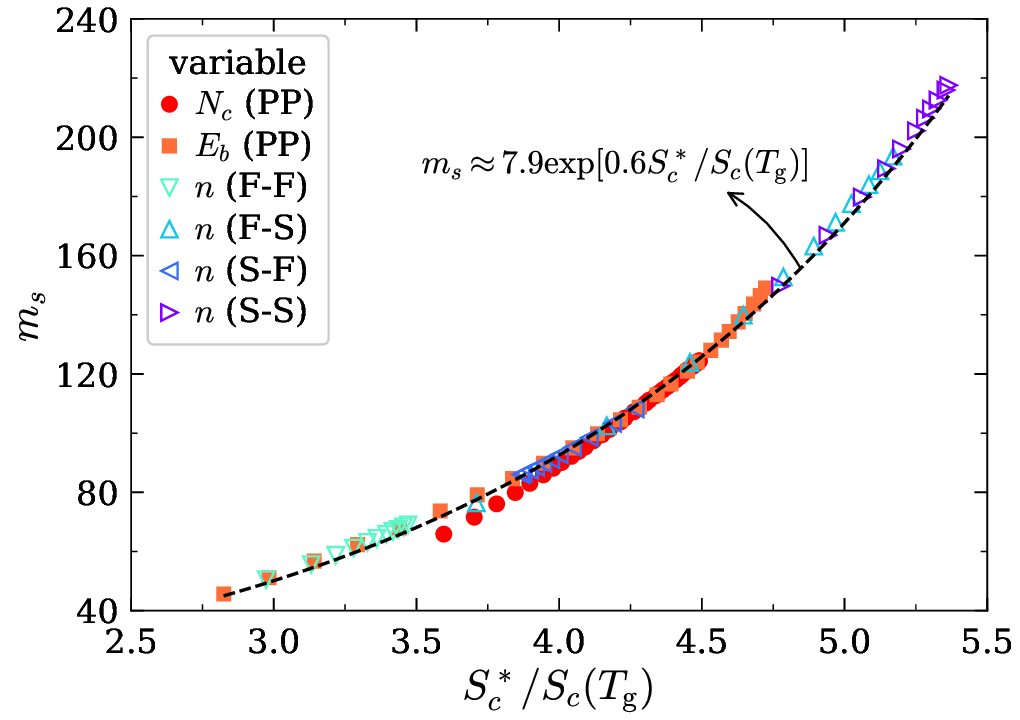}

% Some text to explain the graphic.
	
\end{tocentry}
	
%% if an abstract is not used by the target journal.	%%%%%%%%%%%%%%%%%%%%%%%%%%%%%%%%%%%%%%%%%%%%%%%%%%%%%%%%%%%%%%%%%%%%%
\newpage
	
\begin{abstract}

We utilize the generalized entropy theory (GET) of glass formation to address one of the most singular and least understood properties of polymer glass-forming liquids in comparison to atomic and small molecule liquids---the often relatively high fragility of the polymer dynamics on a segmental scale, $m_s$. Based on this highly predictive framework of both the thermodynamics and segmental dynamics in terms of molecular structure, polymer backbone and side-group rigidities, and intermolecular interaction strength, we first analyze the relation between $m_s$ and the ratio, $S_c^*/ S_c(T_{\mathrm{g}})$, where $S_c$ is the configurational entropy density of the polymer fluid, $S_c^*$ equals $S_c$ at the onset temperature $T_A$ for non-Arrhenius relaxation, and $T_{\mathrm{g}}$ is the glass transition temperature at which the structural relaxation time $\tau_{\alpha}$ equals $100$ s. While the reduced activation energy estimated from an Arrhenius plot (i.e., differential activation energy) normalized by $k_{\mathrm{B}} T_{\mathrm{g}}$ is determined to be not equal to the actual activation energy, we do find that an apparently general nonlinear relation between $m_s$ and $S_c^*/ S_c(T_{\mathrm{g}})$ holds to a good approximation for a large class of polymer models, $m_s \approx 7.9 \exp [0.6S_c^*/ S_c(T_{\mathrm{g}})]$. The predicted ranges of $m_s$ and $S_c^*/ S_c(T_{\mathrm{g}})$ are consistent with experimental estimates for high molecular-mass polymer, oligomeric, small molecule, and atomic glass-forming liquids. In particular, relatively high values of $m_s$ are found for polymers having complex monomer structures and significant chain stiffness. The variation of $m_s$ with molecular mass, chain stiffness, and intermolecular interaction strength can be traced to the variation of $S_c^*$, which is shown to provide a measure of packing frustration defined in terms of the dimensionless thermal expansion coefficient and isothermal compressibility. The often relatively high fragility and large extent of cooperative motion are found in the GET to derive from the often relatively large packing frustration in this class of polymer glass-forming liquids. Finally, we also develop a tentative model of the ``dynamical segmental relaxation time'' based on the GET, in which the polymers on a coarse-grained scale are modeled as strings of structureless ``beads'', as assumed in the Rouse and reptation models of polymer dynamics.

\end{abstract}

\newpage

\section{\label{Sec_Intro}Introduction}

Angell \cite{Relaxation_1991_131-133_13, Formation_1995_267_1924, Entropy_1997_102_171} introduced the concept of ``fragility'' of glass-forming (GF) liquids as an outgrowth of a larger problem of developing a classification scheme for different classes of GF liquids based on the relative strength of the temperature ($T$) dependence of their dynamical properties. In particular, he directly followed the precedent of Laughlin and Uhlmann \cite{Viscous_1972_76_2317} and others \cite{Theoretical_1957_30_8, Structural_2007_353_4613} by considering the logarithm of the shear viscosity $\eta$ as a function of $T_{\mathrm{g}} / T$ with $T_{\mathrm{g}}$ being the experimentally estimated glass transition temperature, where this reduced variable description was motivated by the hypothesis that a corresponding state description might apply generally to GF liquids. \cite{Viscous_1972_76_2317} The consideration of $\log \eta$ in this type of reduced variable plot reflects the common experience that the dynamical properties of liquids (e.g., shear viscosity, structural relaxation time, diffusion coefficient, etc.) generally follow an Arrhenius temperature dependence to a good approximation at relatively high temperatures, an observation that is naturally accommodated by classical transition state theory (TST). \cite{Theory_1941_28_301, Book_Eyring} There is a long history of this type of semi-empirical reasoning in the reduced temperature description of the dynamical properties of fluids. \cite{Viscous_1972_76_2317}

Laughlin and Uhlmann \cite{Viscous_1972_76_2317} observed that while the introduction of the reduced variable description did not lead to a ``universal'' description of $\eta$ for \textit{all} GF liquids, it did seem to apply as a good approximation for specific classes of GF liquids. The important idea was then born of using this type of plot to classify GF liquids. Starting from this exploratory work, Angell \cite{Relaxation_1991_131-133_13, Entropy_1997_102_171} took the next step of introducing a \textit{quantitative measure} of the degree of non-Arrhenius dynamics, which could readily be measured experimentally, as a basis for a precise metrology for classifying GF liquids. In particular, he chose the ``steepness'' $m$, i.e., the local slope, of the Angell plot in the vicinity of $T_{\mathrm{g}}$ as an objective measure of the degree of non-Arrhenius dynamics of any GF liquid. This definition allowed for a more refined categorization of liquids into classes for which this parameter takes similar values. The data base of $m$ values has grown until the present day.

Angell \cite{Relaxation_1991_131-133_13, Entropy_1997_102_171} applied the term ``fragility'' to $m$ in terms of an abstract energy landscape interpretation based on an Adam-Gibbs (AG) \cite{Temperature_1965_43_139} picture of the origin of the temperature dependence of the activation free energy of GF liquids. In the AG model, the non-Arrhenius temperature dependence of relaxation in GF liquids can be traced to the temperature dependence of the configurational entropy $S_c(T)$, a quantity that naturally depends on the structure of the free energy landscapes that ultimately derives from variations in the local intermolecular interactions within the liquid. (We discuss this phenomenon below in further detail, along with relevant references.) This attractive, qualitative, and interpretative view of the variability of the fragility of glass formation in liquids no doubt played a large role in the embrace of fragility in the field of GF liquids as a basic metrical parameter in their classification. In this energy landscape-based terminology, \cite{Relaxation_1991_131-133_13, Entropy_1997_102_171} GF liquids for which the deviations from high temperature Arrhenius relaxation are relatively weak are termed ``strong'', while ``fragile'' GF liquids are those for which the deviations are large. Angell and coworkers \cite{Thermodynamic_1999_398_492, Thermodynamic_2001_410_663} provided substantial experimental evidence supporting this interpretation of fragility in a range of small molecule GF liquids, and Sastry \cite{relationship_2001_409_164} provided evidence in accord with this view in molecular dynamics (MD) simulations of the Kob-Andersen model. However, a later study by Huang and McKenna \cite{New_2001_114_5621} showed that the estimation of fragility derived from the configurational entropy estimated from specific heat measurements did not seem to apply well to polymeric liquids. This discrepancy might possibly reflect the great difficulty of estimating the configurational entropy of polymer liquids, given the especially large vibrational contribution to the entropy of polymer liquids, which has been argued to lead to large uncertainties when this contribution is subtracted from the total entropy obtained from specific heat data to estimate $S_c$. \cite{Relationship_2013_138_12A541, Contributions_2000_112_7518, Thermodynamic_2022_55_8699} This complicated experimental situation regarding the direct interpretation of fragility in terms of $S_c$ has subsequently led to a proliferation of ideas and models \cite{Resolving_2009_102_248301, Heterogeneous_2018_51_2887, Polymer_2001_114_9156, Poisson_2006_86_1567, Revisiting_2019_12_2439, Poisson's_2011_10_823} aimed at theoretically explaining, or at least effectively correlating, fragility with molecular parameters, given the theoretical and practical importance of fragility in classifying GF liquids.

Regardless of the awkward terminology often utilized in discussing ``fragility'', and different points of view about how one might think about this property from a theoretical standpoint, fragility has emerged as a parameter of important engineering significance because it quantifies the relative ``strength'' of the temperature dependence of relaxation in GF liquids, a property highly relevant to the processing and end-use properties of GF materials in diverse applications. Fragility is then important even if one really does not care what this parameter ``means'' theoretically. \cite{Volume_2012_109_185901, Kinetic_2016_29_023002, Structural_2014_5_4616, Compositional_2022_13_3708, Perspective_2023_134_010902, Solvent_2013_9_5336}

It seems inevitable from this confluence of rising theoretical and practical interest in fragility that ever-increasing efforts are being made to understand fragility in terms of molecular parameters \cite{Resolving_2009_102_248301, Heterogeneous_2018_51_2887, Polymer_2001_114_9156, Poisson_2006_86_1567, Revisiting_2019_12_2439, Poisson's_2011_10_823} and explore correlations between fragility and both thermodynamic and dynamic properties of GF liquids that might aid in predicting this fundamental property. At present, there are multiple ``competing'' models of glass formation and corresponding models of the origin of fragility. As we shall discuss below, many of the correlations of fragility with other properties have been repeatedly found to be not as universal as initially hoped. The reason for this unsatisfactory situation can at least partially be attributed to the fact that glass formation and fragility are both not sufficiently understood, so by extension, there is a limited logical basis for predicting how fragility should be related to other properties of GF liquids.

The present work involves a return to Angell’s original energy landscape conception of fragility, but we do not rely on inherent large uncertainties in the experimental estimation of the fluid configurational entropy, $S_c$. In particular, we utilize a well-defined statistical mechanical model \cite{Lattice_1998_103_335, Lattice_2014_141_044909} that allows for the direct computation of $S_c$ and, indeed, essentially any other thermodynamic property of interest, in conjunction with the AG model, \cite{Temperature_1965_43_139} to predict the segmental relaxation time as a function of monomer structure, intermolecular interaction strength, and intramolecular chain rigidity, and general thermodynamic conditions (e.g., temperature and pressure). \cite{Generalized_2008_137_125, Polymer_2021_54_3001, Thermodynamic_2023_41_1329, Advances_2023_53_616} Our calculations support the prescient nature of Angell’s interpretation of the origin of fragility variations in GF liquids.

The present work was initially motivated by the common observation that polymer liquids have a significantly higher fragility than small molecule, atomic, and other non-polymeric GF liquids, but only speculation exists for why this might be the case. \cite{Why_2007_19_205116, Role_2010_43_8977, Why_2016_145_154901} For example, Dalle-Ferrier et al. \cite{Why_2016_145_154901} have argued that ``many polymers cannot reach an ergodic state on the time scale of segmental dynamics due to chain connectivity and rigidity''. Colmenero \cite{Are_2015_27_103101} has gone so far as claim that ``there is a fundamental difference between the nature of the glass transition in polymers and in simple (standard) glass-formers''. It has also been found that correlations between fragility and other properties of GF materials ``break down'' in polymeric GF liquids. \cite{Why_2007_19_205116, Breakdown_2006_39_3322, Temperature_1965_69_3480, Viscoelastic_1971_9_209, Temperature_1980_12_43, Temperature_1982_20_729, Viscoelastic_1995_28_6432} These observations collectively suggest that glass formation might not have a ``universal'' character so that different types of GF liquids would have to be defined and require their own corresponding theories to predict their properties. Alternatively, one might wonder if these deviations from expectations might arise from the limitations of existing theories of glass formation to anticipate the large variation of fragility with molecular structure within a framework in which glass formation in all materials can be viewed as a universal phenomenon. The present work is motivated by this problem of both fundamental and practical interest.

It should also be mentioned that not all polymer GF liquids have a high fragility at the segmental scale, which is another fact that any acceptable theory of polymer glass formation must explain. It is also observed that the oligomer form of polymer fluids (whose high mass counterparts are fragile) tend to have a segmental fragility similar to small molecule and atomic fluids. \cite{Why_2007_19_205116} This is either a clue or a conundrum, depending on how one thinks about the problem. \cite{Why_2007_19_205116} To add to this complexity in understanding fragility, Sokolov and coworkers \cite{Why_2016_145_154901, Breakdown_2006_39_3322} have convincingly demonstrated that the fragility at the scale of the polymer segments can be much higher than the fragility defined at the scales of the entire molecule, which is of importance for understanding the polymer diffusion coefficient and polymer melt shear viscosity. This difference in fragility at different length scales lies at the origin of the breakdown of the time-temperature superposition in many polymers, \cite{Temperature_1965_69_3480, Temperature_1982_20_729, Breakdown_2006_39_3322} among other physical phenomena of great theoretical and practical interest (e.g., the correct modeling of the friction coefficient in coarse-grained models of polymer melts such as the Rouse and reptation models \cite{Viscoelastic_1980_, Identification_1995_68_376}), so fragility of the polymer dynamics at a segmental scale, $m_s$ (defined below), must clearly be distinguished from fragility at the chain scale. \cite{Why_2016_145_154901, Breakdown_2006_39_3322, Surprising_2018_51_4874} The origin of this length scale dependence of the fragility in polymer fluids is another fundamental unresolved problem in polymer melt dynamics, for which there has also been much semi-empirical modeling or correlative-oriented studies. \cite{Resolving_2009_102_248301, Heterogeneous_2018_51_2887, Polymer_2001_114_9156} Although this is not the main topic of the present paper, we tentatively argue for the basis of this dramatic phenomenon within the framework of the generalized entropy theory (GET) \cite{Generalized_2008_137_125, Polymer_2021_54_3001, Thermodynamic_2023_41_1329, Advances_2023_53_616} based on the simple idea that we should view the \textit{same} polymer chains having structured monomers with different chemical interactions, structural shape, and variable stiffness of the backbone and side groups as an effectively flexible chain of simple spherical segments when the chain is viewed in a coarse-grained way on large length scales. (This view seems less radical to a previous successful coarse-graining methodology in which the entire chain is modeled as either a single or a few chain segments. \cite{Coarse-graining_2018_14_7126}) As we shall see below, this simple coarse-graining argument would imply from the GET that all polymers should have significantly lower fragility on the scale of the polymer chains, as observed experimentally. \cite{Why_2007_19_205116, Why_2016_145_154901, Surprising_2018_51_4874}

The present work begins by examining the relation between $m_s$ and the extent of cooperative motion near $T_{\mathrm{g}}$, $S_c^*/ S_c(T_{\mathrm{g}})$, as a function of the molecular mass $N_c$, intermolecular interaction strength $\epsilon$, applied pressure $P$, and rigidities $E_b$ and $E_s$ of the polymer backbone and side groups based on the GET, \cite{Generalized_2008_137_125, Polymer_2021_54_3001, Thermodynamic_2023_41_1329, Advances_2023_53_616} which provides a unique, highly predictive theoretical framework for calculating $m_s$ and other basic thermodynamic and dynamic properties that might be related to $m_s$. The simple reasoning motivating the present work is that if the GET is a valid model of polymer GF liquids, then it should be able to provide a clear explanation of the relatively high segmental fragility of many GF polymers, in addition to explaining why some polymer and oligomeric GF liquids are often found to exhibit a relatively strong glass formation that is similar to many metallic and small molecule GF liquids. By addressing this pointed, but general, question of exactly what molecular factors influence $m_s$, we obtain new insights into the glass formation itself, the extent of cooperative motion in GF liquids and its precise relation with $m_s$, and the role of packing frustration in determining $m_s$. Moreover, the GET should, in principle, allow for the prediction of exactly what molecular parameters and physical attributes of the polymer melt are responsible for the relatively large or small segmental fragility values for the purposes of materials design in which segmental fragility can be highly relevant. \cite{Polyisobutylene_2008_46_1390}

As with other models of glass formation, the GET is based on assumptions whose fundamental validity remains somewhat uncertain. The strengths and weaknesses of the GET have been discussed in previous reviews. \cite{Generalized_2008_137_125, Polymer_2021_54_3001} One particular strength of this model is that it makes unequivocal predictions about both the thermodynamic and dynamic properties of polymer GF liquids within an analytic theoretical framework. As such, the predictions provide useful results for guiding simulation studies to check the validity of the GET, and our group has made considerable progress along this line. \cite{Influence_2016_49_8355, Influence_2016_49_8341, Stringlike_2016_5_1375, Influence_2017_50_2585, Molecular_2020_53_4796, Investigation_2020_53_6828, Understanding_2020_53_7239, Role_2020_53_9678, Equation_2021_54_3247, Influence_2021_54_6327, Influence_2021_54_9587, Melt_2022_55_3221, Thermodynamic_2022_55_8699, Molecular_2023_56_4049, Understanding_2023_41_1447, Parallel_2023_56_4929, Confinement_2024_160_044503} The present work attempts to utilize the GET in a different mode to understand general fundamental aspects of polymer glass formation, such as the relation between $m_s$ and the degree of cooperative motion, the relation between fragility and $T_{\mathrm{g}}$ and other characteristic temperatures of GF liquids, with a particular focus on the problem of why many polymers exhibit a relatively high segmental fragility. The fact that the GET also describes the thermodynamic properties of the same material means that it is possible to identify what thermodynamic properties are germane to understanding the segmental dynamics and the relation between these properties as a function of molecular and thermodynamic parameters.

Our analysis indicates that the high fragility of some polymer GF liquids occurs simply because these fluids have a relatively high packing frustration, but otherwise these fluids are just ``garden-variety'' GF liquids, so it is not necessary to invoke different classes of GF liquids based on molecular structure and chemistry or to invoke ill-defined non-equilibrium mechanisms for understanding their dynamics. The relatively high $m_s$ values of some polymer GF liquids having complex monomer structure and significant backbone or side-group stiffness can be traced to the variation of a previously neglected parameter in the GET, the configurational entropy density of the fluid in the high temperature or ``athermal'' limit, $S_c^*$. It had been possible to calculate $S_c^*$ from the beginning of the formulation of the GET, \cite{Direct_2005_123_111102, Glass_2005_109_21285, Fragility_2005_109_21350, Generalized_2008_137_125} but the physical significance of this parameter was largely ignored in previous works, including more recent papers, \cite{Influence_2014_47_6990, Generalized_2015_48_2333, Generalized_2016_145_234509, Entropy_2016_161_443} simply because it was not clear how to physically think about it. $S_c^*$ provides a measure of the complexity of the free energy landscape defined in terms of the total number of distinct energy minima which defines the configurational entropy. This ``thermodynamic depth'' parameter is found to vary inversely with the efficiency of packing frustration as defined in the GET by the dimensionless thermal expansion coefficient and isothermal compressibility. \cite{Polymer_2021_54_3001, Entropy_2016_161_443} Therefore, the often relatively high fragility and large extent of cooperative motion are found in the GET to derive from the often relatively large packing frustration in this class of polymer GF liquids. Finally, we develop a tentative model of the ``dynamical segmental relaxation time'' based on the GET, in which the polymers on a coarse-grained scale are modeled as strings of structureless ``beads'', as assumed in the Rouse and reptation models of polymer dynamics. \cite{Viscoelastic_1980_}

\section{\label{Sec_GET}Generalized Entropy Theory}

The GET provides a powerful predictive theoretical framework for investigating the thermodynamics of polymer fluids and their nontrivial segmental dynamics associated with glass formation. \cite{Generalized_2008_137_125, Polymer_2021_54_3001, Thermodynamic_2023_41_1329, Advances_2023_53_616} This theory involves the combination of the analytic lattice cluster theory (LCT), \cite{Lattice_1998_103_335, Lattice_2014_141_044909} which allows for systematic calculations of the configurational entropy $S_c$ of polymer melts having general monomer structure, variable chain rigidity, the presence of side groups, and intermolecular interactions governing the intermolecular cohesive interaction strength of the polymer fluid, and for general thermodynamic conditions (e.g., temperature $T$, pressure $P$, etc.), and the AG model \cite{Temperature_1965_43_139} relating the structural relaxation time $\tau_{\alpha}$ to $S_c$. While the GET evidently emphasizes the thermodynamic aspects of polymer glass formation, there have been attempts to relate the configurational entropy to the structural information of GF liquids. \cite{Atomistic_2020_101_014113, Role_2014_113_225701, Structural_2012_137_024508} In particular, Han et al. \cite{Atomistic_2020_101_014113} have shown that the distribution of local Voronoi polyhedra is related to the configurational entropy based on metallic glass-forming materials, consistent with the view that the configurational entropy contains higher-order structural correlations beyond the pair correlation level.

The GET predicts $\tau_{\alpha}$ to take the following general form,
\begin{eqnarray}
	\label{Eq_AG}
	\tau_{\alpha} = \tau_o \exp \left[ \frac{\Delta G_o}{k_{\mathrm{B}} T} \frac{S_c^*}{S_c(T)} \right]
\end{eqnarray}
where $k_{\mathrm{B}}$ is Boltzmann's constant, $S_c(T)$ is the configurational entropy density, and $S_c^*$ is the maximum of $S_c(T)$ and is naturally defined at the onset temperature $T_A$ for non-Arrhenius relaxation. AG originally interpreted $S_c^*/ S_c(T_{\mathrm{g}})$ as defining the extent of cooperative motion or the ``size'' of the cooperatively rearranging regions (CRRs) in cooled liquids in terms of the number of molecular units involved in the motion rather than the geometrical size, \cite{Temperature_1965_43_139} but we do not need to invoke this interpretation in our theoretical development. However, we note that previous MD simulations have shown the existence of string-like collective motion whose variation with temperature is consistent with the assumptions of AG. \cite{Stringlike_1998_80_2338, Relationship_2013_138_12A541, Does_2006_125_144907, Quantitative_2015_112_2966, String_2014_140_204509} The high temperature vibrational prefactor $\tau_o$ is expected to range from $10^{-14}$ s to $10^{-13}$ s, \cite{Universality_2003_67_031507, Temperature_2015_48_3005} but we adopt a constant value of $\tau_o = 10^{-13}$ s in the present paper, following the original formulation of the theory. \cite{Generalized_2008_137_125} For any specific fluid at equilibrium, $\tau_o$ can be estimated precisely from the initial decay of the velocity autocorrelation function (see Section D of the Supplementary Information in ref \citenum{Fast_2021_154_084505} for an illustrative estimation of $\tau_o$ in the case of a metallic glass material). $\Delta G_o$ is the activation free energy at high $T$, which is anticipated from classical TST \cite{Theory_1941_28_301, Book_Eyring} to contain both enthalpic $\Delta H_o$ and entropic $\Delta S_o$ contributions, i.e., $\Delta G_o = \Delta H_o - T \Delta S_o$. Motivated by the heuristic approximation made by AG \cite{Temperature_1965_43_139} that the entropic contribution to the activation free energy is negligible, i.e., $\Delta S_o = 0$, along with the simulation evidence for a limited number of coarse-grained liquids and some experimental evidence for GF liquids indicating the approximation, \cite{Generalized_2008_137_125} $\Delta H_o \approx (7 \pm 1) \ k_{\mathrm{B}} T_c$, the original GET model simply assumed that $\Delta G_o \approx \Delta H_o \approx 6 \ k_{\mathrm{B}} T_c$ for a rough estimation of the structural relaxation time of polymers having complex structure with no other parameters other than those required to describe the thermodynamics of the polymer material. \cite{Generalized_2008_137_125} Here, $T_c$ is the ``crossover temperature'' in the GET that separates the high- and low-$T$ regimes of glass formation and which is precisely defined theoretically by an inflection point in $TS_c(T)$. \cite{Generalized_2008_137_125}

\begin{figure*}[htb]
	\centering
	\includegraphics[angle=0,width=0.45\textwidth]{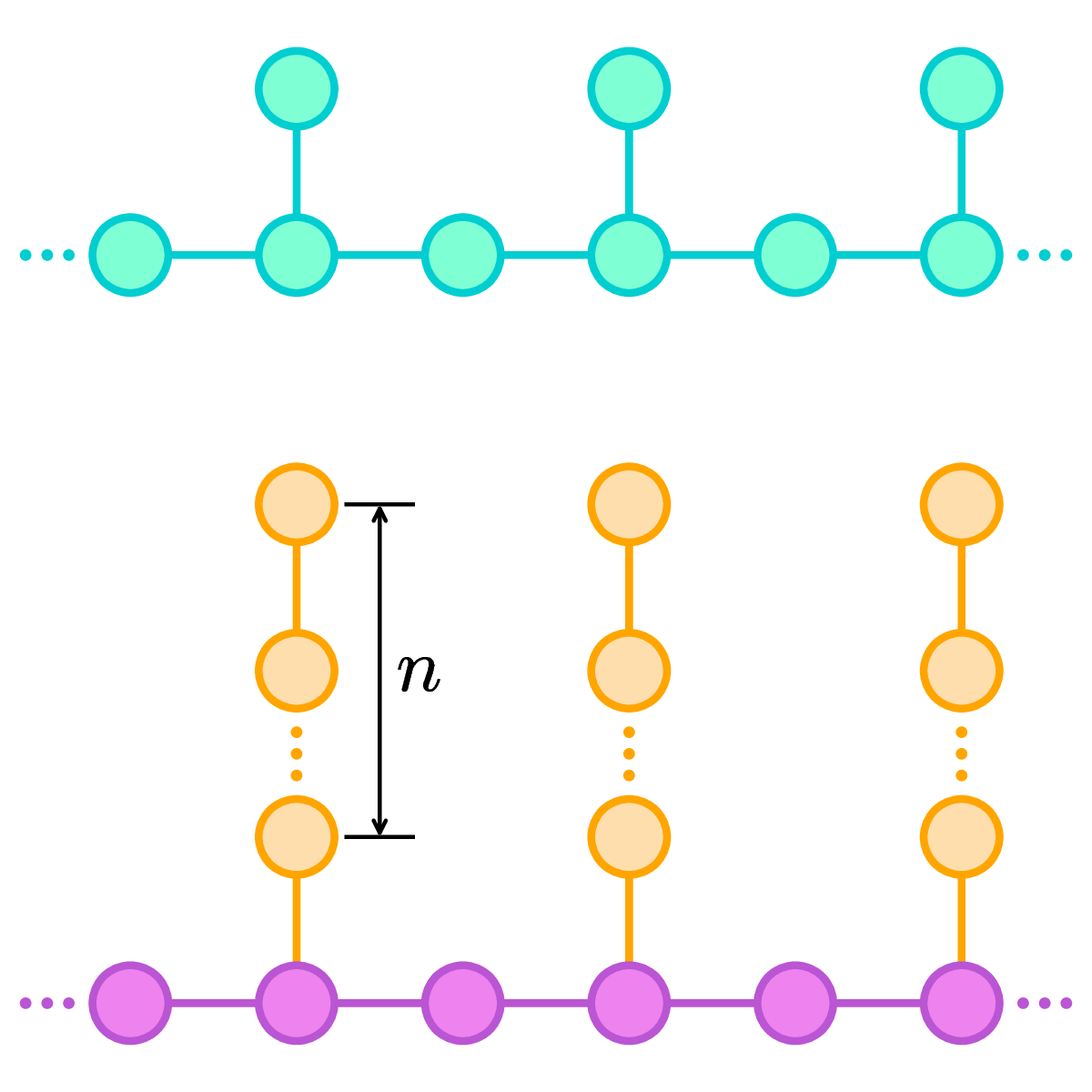}
	\caption{Cartoons of representative polymer models in the generalized entropy theory (GET). Circles designate united-atom groups, solid lines depict the bonds between united-atom groups, and dotted lines indicate the bonds linking the united-atom groups to their neighbors along the chains. The upper part corresponds to the structure of polypropylene (PP), for which a single bending energy parameter $E_b$ is adequate. The lower part corresponds to the structure of poly($n$-$\alpha$-olefins), where different colors indicate that separate chain stiffness parameters can be assigned to the backbone and groups. Each monomer is composed of a set of united-atom groups. The basic molecular parameters in the lattice model include the number $N_c$ of repeating monomers in an individual chain, the length $n$ of side groups, the cohesive energy parameter $\epsilon$ between nearest-neighbor united-atom groups, and the bending energy parameters $E_b$ and $E_s$ for the backbone and side groups having two or more united-atom groups.}\label{Fig_Scheme}
\end{figure*}

\begin{table}
	\caption{Bending energy parameters $E_b$ and $E_s$ of the backbone and side groups for the flexible-flexible (F-F), flexible-stiff (F-S), stiff-flexible (S-F), and stiff-stiff (S-S) classes of polymers in the GET.}\label{Table_Class}
	\begin{tabular}{lcc}
		\hline
		polymer class & $E_b / k_{\mathrm{B}}\ [\mathrm{K}]$ & $E_s / k_{\mathrm{B}}\ [\mathrm{K}]$ \\
		\hline
		F-F & $200$ & $200$ \\
		F-S & $200$ & $600$ \\
		S-F & $600$ & $200$ \\
		S-S & $600$ & $600$ \\
		\hline
	\end{tabular}
\end{table}

To aid in the discussion below, Figure \ref{Fig_Scheme} shows representative polymer models in the LCT. The numbers of repeating monomers and united-atom groups in each side group are denoted by $N_c$ and $n$, respectively, so the molecular mass is given by $M = N_c (n + 2)$. For a pair of consecutive bonds, the LCT treats the conformational energy difference between \textit{trans} and \textit{gauche} conformations in terms of the bending energy parameter $E_b$. To better account for chemical diversity in monomer structures, two separate bending energy parameters, $E_b$ and $E_s$, can be assigned to the backbone and side groups, \cite{Generalized_2008_137_125} respectively. This feature is highlighted by different colors for the backbone and side groups of the polymer model shown in the lower part of Figure \ref{Fig_Scheme}. Moreover, the LCT permits prescribing three different cohesive energy or well-depth parameters for polymers having two types of united-atom groups in the monomer. \cite{Lattice_2014_141_044909, Generalized_2015_48_2333} For simplicity, however, the present work focuses on the case of a common monomer-averaged cohesive energy parameter $\epsilon$. Accordingly, the LCT allows for a classification of polymers based on the relative values of $E_b$ and $E_s$, inspired by the work of Colucci and McKenna, \cite{Fragility_1997_455_171} who studied correlative relations between molecular structure and fragility for a range of polymers whose monomer structures were schematically illustrated. More specifically, the flexible-flexible (F-F), flexible-stiff (F-S), stiff-flexible (S-F), and stiff-stiff (S-S) classes of polymers correspond to chains with a flexible backbone and flexible side groups, chains with a flexible backbone and relatively rigid side branches, chains with a relatively stiff backbone and flexible side groups, and chains with both a stiff backbone and stiff side groups, respectively. \cite{Polymer_2021_54_3001, Understanding_2020_53_7239, Influence_2021_54_6327} In most cases, we set the values of $E_b$ and $E_s$ in Table \ref{Table_Class}, but we use a different set of $E_b$ and $E_s$ than those in Table \ref{Table_Class} when we develop the model of chain dynamics based on the GET in Section \ref{Sec_Length}.

\begin{table}
	\caption{Molecular and thermodynamic parameters utilized in the calculations based on the GET for the structure of polypropylene (PP) having different variables.}\label{Table_Para}
	\begin{tabular}{lcccc}
		\hline
		variable & $N_c$ & $E_b / k_{\mathrm{B}}\ [\mathrm{K}]$ & $\epsilon / k_{\mathrm{B}}\ [\mathrm{K}]$ & $P\ [\mathrm{MPa}]$\\
		\hline
		chain length & $6$--$8000$ & $600$ & $200$ & $0.101325$ \\
		chain rigidity & $8000$ & $200$--$800$ & $200$ & $0.101325$ \\
		cohesive energy & $8000$ & $600$ & $100$--$300$ & $0.101325$ \\
		pressure & $8000$ & $600$ & $200$ & $0$--$100$ \\
		\hline
	\end{tabular}
\end{table}

In all our calculations below, the lattice coordination number is fixed at $z = 6$ and the cell volume parameter is fixed at $V_{\mathrm{cell}} = 0.25^3 \ \mathrm{nm}^3$. Our calculations consider both the structure of polypropylene (PP) and the different classes of polymers, as shown in Figure \ref{Fig_Scheme}. For the PP structure, a single bending energy parameter $E_b$ for the backbone is adequate, and we vary $N_c$, $E_b$, $\epsilon$, and $P$ individually and the specific parameters are listed in Table \ref{Table_Para}. For different classes of polymers, we focus on the variable $n$, which is varied from $2$ to $13$, and the remaining parameters are set as follows, $M = 24000$, $\epsilon / k_{\mathrm{B}} = 200$ K, and $P = 0.101325$ MPa. Note that the magnitude of $\epsilon / k_{\mathrm{B}} = 200$ K is typical for polyolefins. \cite{Generalized_2008_137_125}

We mention that numerous MD simulation studies have confirmed the existence of cooperative exchange motion in different types of GF liquids, \cite{Stringlike_1998_80_2338, Relationship_2013_138_12A541, Does_2006_125_144907} consistent with the hypothesis of such collective particle exchange motion by AG. \cite{Temperature_1965_43_139} Freed \cite{Communication_2014_141_141102} has offered an extension of TST that rationalizes these simulation observations. The abstract CRRs of AG have been found to take the geometrical form of polymeric structures that dynamically form and disintegrate. \cite{Stringlike_1998_80_2338, Relationship_2013_138_12A541, Does_2006_125_144907} These simulation observations stimulated the development of the string model of glass formation based on the observed properties of these string-like dynamical structures. \cite{Communication_2014_141_141102, String_2014_140_204509, Polymer_2021_54_3001} Although we do not discuss this model in our development below, we simply note that $S_c^*/ S_c(T_{\mathrm{g}})$ equals the magnitude of the average string length $L(T)$ divided by its value at $T_A$ in the string model of glass formation. \cite{String_2014_140_204509} An important aspect is that the string model additionally treats the thermodynamics of string formation in terms of an equilibrium polymerization model, \cite{String_2014_140_204509, Does_2006_125_144907} which allows for a precise modeling of the observed dynamic heterogeneities and their impact on the fluid dynamics, thereby taking the entropy models to a more molecular level of description and enabling more quantitative tests of the entropy theory of glass formation. Our main point in this discussion is that there is ongoing work to provide a firmer foundation for the GET, which was admittedly based more on ``inspiration'' and observation rather than an actual theoretical derivation.

\section{Results and Discussion}

\subsection{\label{Sec_Fragility}Direct Computation of Segmental Fragility and Extent of Cooperative Motion}

Many experimental and computational studies have followed the lead of Angell \cite{Relaxation_1991_131-133_13, Formation_1995_267_1924, Entropy_1997_102_171} and have classified GF liquids as being ``strong'' or ``fragile'' based on whether the material has a dynamics that is nearly Arrhenius or strongly deviates from Arrhenius behavior, respectively. Various measures of the ``degree'' of fragility have been introduced to more quantitatively estimate the degree of deviation from Arrhenius dynamics. This term does not refer to material strength and corresponds to the engineering terms ``long'' and ``short'' \cite{Poisson_2006_86_1567, Viscous_1972_76_2317} derived from the working time of a material in fabrication processes in which the material cools at more or less constant rate.

The most common quantitative measure of fragility in experimental studies is the ``steepness index parameter'' $m$, \cite{Relaxation_1991_131-133_13, Formation_1995_267_1924, Entropy_1997_102_171}
\begin{equation}
	\label{Eq_FraDefinition}
	m = \left. \frac{\partial \log\tau_{\alpha}}{\partial (T_{\mathrm{g}} / T)} \right|_{P, T = T_{\mathrm{g}}}
\end{equation}
which describes the local slope in an Arrhenius plot of the structural relaxation time $\tau_{\alpha}$ or other relaxation times from stress or dielectric relaxation or some other dynamical property (e.g., diffusion coefficient and shear viscosity) at $T_{\mathrm{g}}$, divided by $k_{\mathrm{B}} T_{\mathrm{g}}$ to make the parameter dimensionless. Estimating $m$ and correlating this parameter with other properties that can be measured represent a kind of ``industry'' in the field of GF liquids with many computational and experimental participants.

As noted in Section \ref{Sec_Intro}, the magnitude of the fragility at the segmental scale in polymer liquids or ``segmental fragility'' $m_s$ tends to much larger in polymer fluids than atomic and small molecule GF liquids. \cite{Why_2007_19_205116, Role_2010_43_8977, Why_2016_145_154901} This significant difference raises questions about the universality of $m_s$ in different fluids in relation to other properties of GF liquids that otherwise appear quite similar to other types of GF liquids. To address this widely observed experimental trend, we directly calculate $m_s$ from the GET based on the \textit{exact} definition of this property in this model,
\begin{equation}
	\label{Eq_FraGET}
	m_s = \frac{\Delta H_o}{k_{\mathrm{B}} \ln 10} \left[ \frac{S_c^*}{S_c(T_{\mathrm{g}})} \right]^2 \left. \frac{\partial \{(T / T_{\mathrm{g}}) [S_c(T_{\mathrm{g}}) / S_c^*]\}}{\partial T} \right|_{P, T=T_{\mathrm{g}}}
\end{equation}
where the configurational entropy density $S_c(T_{\mathrm{g}})$ at $T_{\mathrm{g}}$ of importance for materials under constant pressure and highly variable temperature conditions is reduced by its high temperature or ``athermal'' limit $S_c^*$ and the reduced temperature is defined by $T / T_{\mathrm{g}}$. In the GET, $S_c^*$ is defined precisely as a material specific quantity, \cite{Generalized_2008_137_125, Entropy_2016_161_443, Resolution_2000_112_8958} $S_c^* \equiv S_c(T_A)$, rather than taken to be an adjustable parameter, as in the original AG model. \cite{Temperature_1965_43_139} Following the common convention in experimental studies, $T_{\mathrm{g}}$ is precisely defined to be the temperature at which $\tau_{\alpha}$ equals $100$ s. We also note that the string model of glass formation, \cite{String_2014_140_204509, Polymer_2021_54_3001} which took advantage of computational and experimental observations noted in Section \ref{Sec_GET} that the string-like cooperative motion provides a quantitative realization of the hypothetical CRRs of AG, likewise predicts that $m_s$ is related to both the CRR size and its temperature derivative at $T_{\mathrm{g}}$. \cite{Fragility_2013_9_241} However, the near universal relation between $m_s$ and $S_c^* / S_c(T_{\mathrm{g}})$ that we determine is not obvious from the GET and eq \ref{Eq_FraGET}.

In the literature on GF liquids, $m_s$ is often interpreted as the ``apparent activation energy'' at $T_{\mathrm{g}}$ (the local slope in the Arrhenius plot of $\tau_{\alpha}$ or some other dynamical property), divided by the thermal energy $k_{\mathrm{B}} T_{\mathrm{g}}$ to make this parameter dimensionless. However, the \textit{actual} temperature-dependent activation energy $\Delta E_A (T_{\mathrm{g}})$ normalized by $k_{\mathrm{B}} T_{\mathrm{g}}$ in both the AG and GET models is exactly defined by the relation,
\begin{eqnarray}
	\label{Eq_EATg}
	\frac{\Delta E_A(T_{\mathrm{g}})}{k_{\mathrm{B}} T_{\mathrm{g}}} = \frac{\Delta H_o}{k_{\mathrm{B}} T_{\mathrm{g}}} \frac{S_c^*}{S_c(T_{\mathrm{g}})}
\end{eqnarray}

As discussed in Section \ref{Sec_GET}, the AG and GET models neglect the entropy $\Delta S_o$ of activation in the TST expression for $\tau_{\alpha}$, so AG in effect assumed that the activation free energy $\Delta G(T)$ is approximately equal to $\Delta E_A(T)$. Notably, AG neglected $\Delta S_o$ as a matter of mathematical expediency, based on the opinion stated in a footnote, that this term should be relatively ``small'' and thus might reasonably be neglected. \cite{Temperature_1965_43_139} Only recently has this questionable assumption been critically examined and a finite $\Delta S_o$ incorporated into both the entropy and string models of glass formation. \cite{Polymer_2021_54_3001} However, the neglect of $\Delta S_o$ certainly simplifies the determination of $m_s$, and in the present work, we adopt this conventional assumption, which preserves essential trends in the predictions of the GET model based on our experience. We next make some general observations about these ``activation energy'' definitions that are relevant to defining and understanding the physical meaning of $m_s$. Most importantly, it should be appreciated that these activation energy parameters are not at all equivalent!

\begin{figure*}[htb]
	\centering
	\includegraphics[angle=0, width=0.45\textwidth]{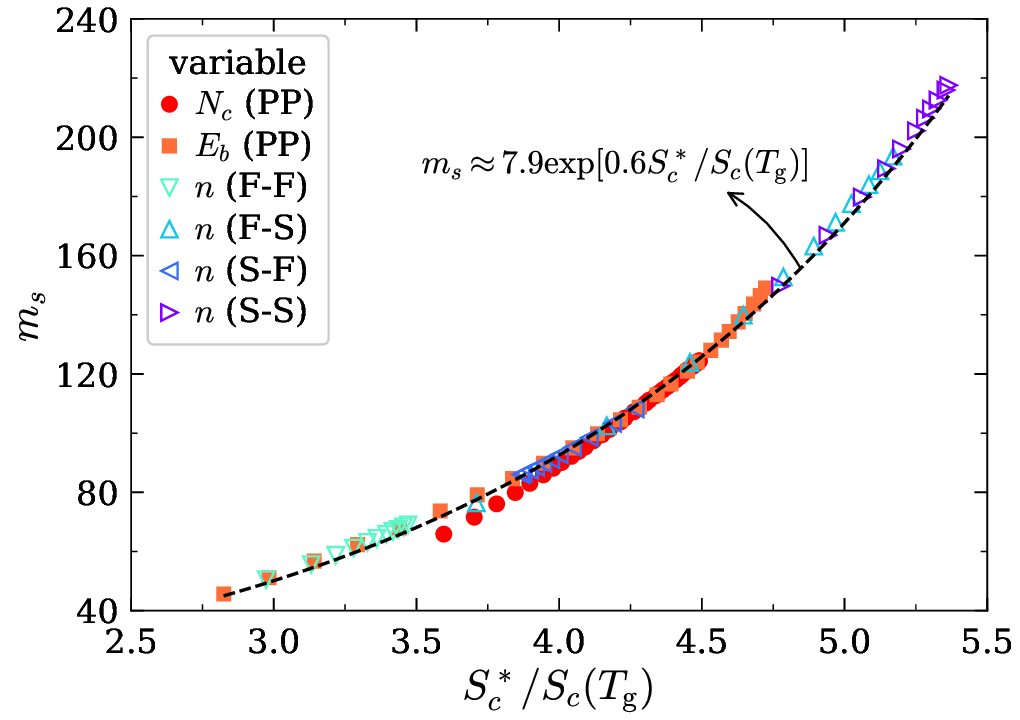}
	\caption{Variation of the segmental fragility $m_s$ as a function of the extent of cooperative motion at the glass transition temperature $T_{\mathrm{g}}$, $S_c^* / S_c(T_{\mathrm{g}})$, calculated from the GET for the PP structure having variable $N_c$ and $E_b$ and for different classes of polymers having variable $n$. The line corresponds to eq \ref{Eq_MZTg}, as indicated in the figure. Note that typical ``flexible'' polymers having flexible side groups (F-F model) have $m_s$ values of typical small molecule liquids, while polymers having stiff backbone or side groups or both (F-S, S-F, and S-S polymer classes) tend to have much larger $m_s$. Increasing the overall chain stiffness ($E_b$) and chain length ($N_c$) causes $m_s$ to vary between these high and low fragility regimes (see circle and square symbols).}\label{Fig_CRR_M}
\end{figure*}

In both the AG \cite{Temperature_1965_43_139} and GET \cite{Generalized_2008_137_125, Polymer_2021_54_3001} models, the change of the activation energy relative to the activation energy in the high temperature regime is directly related to the change in the extent of collective motion precisely defined by $S_c^* / S_c(T)$ (see eq \ref{Eq_EATg}). Thus, one might expect $m_s$ to have \textit{some} relation to this ratio since this quantity is the differential activation energy divided by $k_{\mathrm{B}} T_{\mathrm{g}}$ and since $T_{\mathrm{g}}$ is often found to scale linearly with the high temperature activation energy to a good approximation. Evidently, eq \ref{Eq_FraGET} indicates that the differential activation energy implies a much more complicated quantity than a proportionality between $m_s$ and $S_c^* / S_c(T)$. We then calculate $m_s$ and $S_c^* / S_c(T_{\mathrm{g}})$ for a wide range of polymer models described in Section \ref{Sec_GET} to assess the quantitative nature of this relationship implied by the GET. We show the observed interrelationship for a range of diverse polymers in Figure \ref{Fig_CRR_M}, where we find that $m_s$ increases nearly exponentially with $S_c^* / S_c(T_{\mathrm{g}})$, 
\begin{eqnarray}
	\label{Eq_MZTg}
	m_s \approx 7.9 \exp[0.6S_c^*/ S_c(T_{\mathrm{g}})]
\end{eqnarray}
The analysis in Figure \ref{Fig_CRR_M} includes a PP model having a range of molecular masses and chain rigidities as well as the different classes of polymer melts having variable side-group length $n$. This predicted relation, which we expect to apply to GF liquids more broadly based on the hypothesis of ``universality'', is apparently a new result. (For variable $\epsilon$ and $P$, however, the predictions deviate from the the near universal relation described by eq \ref{Eq_MZTg}, as shown in Section S1 of the Supporting Information, and we expand our discussion under these conditions in Section \ref{Sec_Inverted}.) We note that, in the limit in which glass formation becomes perfectly strong so that $S_c^*/ S_c(T_{\mathrm{g}})$ formally equals $1$, the estimate of $m_s$ from eq \ref{Eq_MZTg} then reduces to a minimum value of about $15$. We may understand this minimum value of $m_s$ to arise from the tendency of $T_0$ in the Vogel-Fulcher-Tammann (VFT) equation for $\tau_{\alpha}$ (see eq \ref{Eq_VFT} below) to approach $0$ as molecular parameters are varied so that the VFT relation reduces exactly to the Arrhenius relation of a perfectly strong GF liquid. In such a case, $m$ based on the VFT equation exactly equals, $m_{\mathrm{min}} = \log [\tau_{\alpha} (T_{\mathrm{g}}) / \tau_{0, \mathrm{VFT}}]$, which takes a value in the range $15$ to $16$ if $\tau_{\alpha} (T_{\mathrm{g}})$ is taken to be $100$ s and $\tau_{0, \mathrm{VFT}}$ is taken to be on the order of $10^{-14}$ s to $10^{-13}$ s (see Section S2 of the Supporting Information). This estimate of $m_{\mathrm{min}}$ is consistent with the empirical estimate of $m = 16$, which has been claimed to be ``established'' from previous experimental studies of GF liquids. \cite{Correlation_2007_19_116103, Nonexponential_1993_99_4201}

The segmental fragility parameter can also be defined and measured for materials under constant density conditions and can readily be calculated within the GET. \cite{Thermodynamic_2013_138_234501} This quantity $m_{_V}$ has been tabulated for numerous polymeric and non-polymeric GF liquids, \cite{Why_2005_72_031503} where it was found that $m_{_V}$ scales linearly with $m_s$ to a good approximation. Direct calculations of $m_{_V}$ versus $m_s$ based on the GET, where $m_s$ is calculated at $0.101325$ MPa ($1$ atm in non-SI units), likewise indicate an approximately linear relation for the same PP polymer model illustrated in Figure \ref{Fig_CRR_M}, where $N_c$, $E_b$, and $\epsilon$ are varied. The GET also indicates that the thermodynamic scaling exponent $\gamma$ varies nearly linearly with the reciprocal of $m_{_V}$, as observed in the experiments of Casalini and Roland. \cite{Why_2005_72_031503} The relation between fragility and the extent of collective motion indicated in Figure \ref{Fig_CRR_M} would seem to imply approximate relationships between $m_{_V}$ and $\gamma$ and the extent of collective motion. Below, our focus is placed on segmental fragility $m_s$ determined under constant pressure conditions taken to be $1$ atm.

\begin{figure*}[htb]
	\centering
	\includegraphics[angle=0, width=0.45\textwidth]{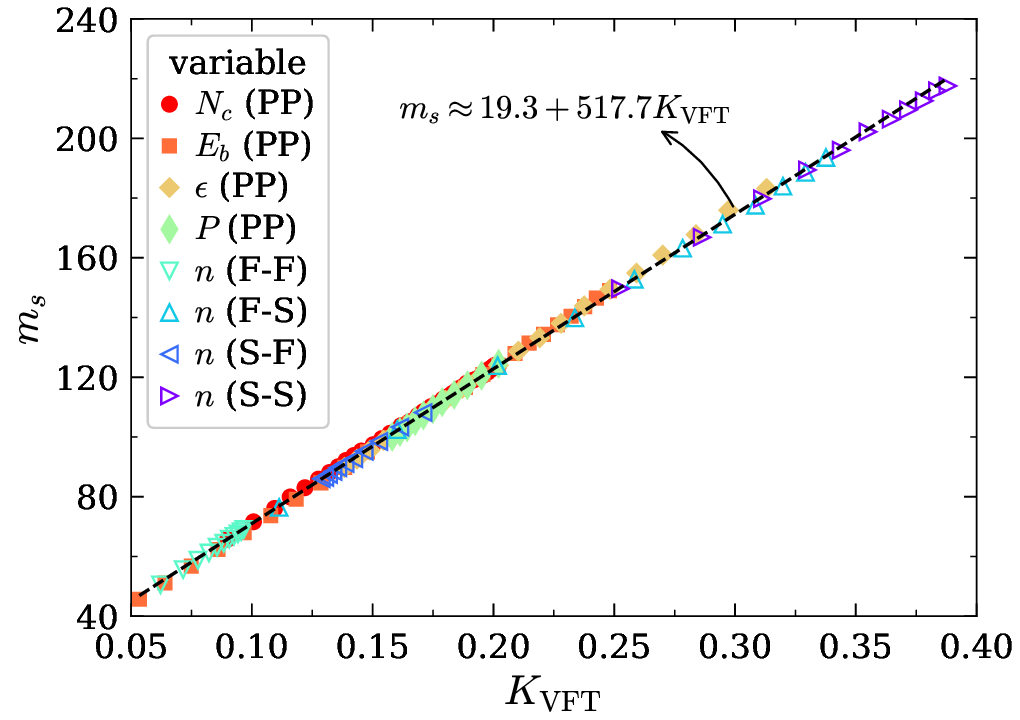}
	\caption{Segmental fragility $m_s$ as a function of the VFT fragility parameter $K_{\mathrm{VFT}}$ calculated from the GET for the PP structure having variable $N_c$, $E_b$, $\epsilon$, and $P$ and for different classes of polymers having variable $n$. The line corresponds to eq \ref{Eq_MKVFT}, as indicated in the figure.}\label{Fig_KVFT_M}
\end{figure*}

There are alternative definitions of fragility in use. In addition to $m_s$, fragility is often defined in therms of the parameter $K_{\mathrm{VFT}}$ estimated from a fit of relaxation time data to the VFT equation, \cite{law_1921_22_645, Analysis_1925_8_339, Abhangigkeit_1926_156_245}
\begin{equation}
	\label{Eq_VFT}
	\tau_{\alpha} = \tau_{0, \mathrm{VFT}} \exp \left[\frac{T_0}{K_{\mathrm{VFT}}(T-T_0)}\right]
\end{equation}
where $\tau_{0, \mathrm{VFT}}$ is a prefactor, $K_{\mathrm{VFT}}$ is the VFT ``fragility index'' that quantifies the degree to which the $T$ dependence of $\tau_{\alpha}$ deviates from an Arrhenius function, and $T_0$ is the temperature at which $\tau_{\alpha}$ formally extrapolates to infinity. ($K_{\mathrm{VFT}}$ is the reciprocal of the fragility strength parameter $D$, i.e., $K_{\mathrm{VFT}} \equiv 1 / D$, so that an increase of $K_{\mathrm{VFT}}$ corresponds to an increase in fragility.) Consistent with many measurement studies, the GET predicts that $\tau_{\alpha}$ can be described by the VFT equation to a high level of approximation in the temperature regime between $T_c$ and $T_{\mathrm{g}}$, where all parameters in this equation are specified by molecular parameters and thermodynamic conditions. Notably, the normally empirical prefactor $\tau_{0, \mathrm{VFT}}$ in eq \ref{Eq_VFT} can directly be calculated using the GET as a function of molecular and thermodynamic parameters and its variation for the polymers considered in the present paper is illustrated in Section S2 of the Supporting Information, where it is shown that $\tau_{0, \mathrm{VFT}}$ has the same order of magnitude as $\tau_o$. We see from Figure \ref{Fig_KVFT_M} that $m_s$ varies linearly with $K_{\mathrm{VFT}}$ predicted from the GET to a high degree of approximation for all the polymer models that we investigate,
\begin{eqnarray}
	\label{Eq_MKVFT}
	m_s \approx 19.3 + 517.7 K_{\mathrm{VFT}}
\end{eqnarray}
This linear relation between $m_s$ and $K_{\mathrm{VFT}}$ is not surprising since the VFT equation implies the following equation,
\begin{eqnarray}
	\label{Eq_VFTRelation}
	m_s = \frac{[\ln \tau_{\alpha}(T_{\mathrm{g}}) - \ln \tau_{0, \mathrm{VFT}}]^2}{\ln10} K_{\mathrm{VFT}} + \frac{\ln \tau_{\alpha}(T_{\mathrm{g}}) - \ln \tau_{0, \mathrm{VFT}}}{\ln 10}
\end{eqnarray}
It is evident that $m_s$ varies with $K_{\mathrm{VFT}}$ in a linear fashion. The GET also allows for the direct computation of the fit parameter $\tau_{0, \mathrm{VFT}}$, as discussed in Section S2 of the Supporting Information.

Measurements and simulations seem to indicate that the ratio of the activation energy from its value in the Arrhenius regime to its value at $T_{\mathrm{g}}$, corresponding to $S_c^* / S_c(T_{\mathrm{g}})$ in eq \ref{Eq_EATg}, is typically around $4 \pm 1$ in polymeric GF liquids. Somewhat smaller values of $m_s$ seem to be characteristic of small-molecule liquids so that the extent of collective motion is broadly consistent with the range of the GET predictions for $S_c^* / S_c(T_{\mathrm{g}})$ indicated in Figure \ref{Fig_CRR_M}. We should also mention a valiant attempt to estimate $S_c^* / S_c(T_{\mathrm{g}})$ by Yamamuro and coworkers \cite{Calorimetric_1998_102_1605, Thermodynamic_2012_109_045701} based on specific heat measurements, where values of $S_c^* / S_c(T_{\mathrm{g}})$ in the somewhat larger range of about $6 \pm 2.5$ were inferred, although these thermodynamic estimates are based on some optimistic assumptions. \cite{Resolution_2000_112_8958} Although the uncertainty is large for these estimates of $S_c^* / S_c(T_{\mathrm{g}})$, the general order of magnitude can be taken as being reliable. Evidently, the variations of $S_c^* / S_c(T_{\mathrm{g}})$ predicted by the GET are not terribly large and, moreover, $S_c^* / S_c(T_{\mathrm{g}})$ is not highly variable between different types of GF liquids. These values are also quite distinct for the local steepness of the Arrhenius plot from which $m_s$ is determined when the glass formation is ``fragile''. We also observe that the range of $S_c^* / S_c(T_{\mathrm{g}})$ corresponds to a strikingly large range of fragility values, $m_s$, that encompass the entire range normally observed in polymer and other diverse GF materials. Specifically, the GET predicts much lower values of $m_s$ under certain conditions that are similar in magnitude to typical values found in metallic and small molecule GF liquids and oligomeric and flexible polymers. In conclusion, the GET predictions naturally encompass the full range of observation of fragility variations in polymeric GF liquids within a single theoretical framework. Extensive tabulations of $m_s$ \cite{Poisson_2006_86_1567, Revisiting_2019_12_2439, Poisson's_2011_10_823} have become available in association with testing a proposed relation between $m_s$ and the material Poisson ratio. \cite{Poisson_2004_431_961} McKenna and coworkers \cite{New_2001_114_5621, Correlation_2006_352_2977} have also provided highly useful tabulations and discussion of fragility values with an emphasis on comparing polymeric to non-polymeric materials.

Later work indicated that the relationship proposed earlier by Novikov and Sokolov \cite{Poisson_2004_431_961} for the relation between fragility and the Poisson ratio simply did not apply to polymers, \cite{Structural_2006_73_064202} but interest in the general idea that fragility might be related to the Poisson ratio remains high, even though the predictive value of correlative relationships is apparently limited to fixed classes of materials. Along the same line, Dalle-Ferrier et al. \cite{Why_2016_145_154901} have provided a ``litany'' of properties that have been first experimentally correlated with $m_s$, along with a list of later experimental studies showing the breakdown of these empirical correlations in the case of polymer materials. The problem of understanding the often relatively high fragility of polymeric GF liquids, and perhaps many other properties of GF liquids, then revolves around an understanding of exactly what the physical factors make $m_s$ large in this class of liquids. We adopt the GET as a computational tool to illuminate this problem below.

\subsection{\label{Sec_STA}$S_c^*$ as a Key Parameter in Glass-Forming Liquids}

It is evident from the predicted relationship between the extent of collective motion in the GET and the reduced configurational entropy, i.e., $S_c^* / S_c(T)$, that the configurational entropy $S_c^*$ at elevated temperatures in which the fluid exhibits Arrhenius dynamics is a significant thermodynamic property in relation to glass formation through its influence on the activation energy $\Delta E_A (T)$ for relaxation in eq \ref{Eq_EATg} and the segmental fragility $m_s$ in eq \ref{Eq_MZTg}.

Despite the evident importance of $S_c^*$ in their theory, AG \cite{Temperature_1965_43_139} could only speculate about its magnitude and they were forced to treat this quantity as an empirical parameter. Johari \cite{Resolution_2000_112_8958} later emphasized the importance of $S_c^*$ as a \textit{fundamental material parameter}, but he was not able to offer any insight into how this quantity might be estimated in terms of molecular parameters and thermodynamic conditions. One of the advantages of the marriage of the LCT for the thermodynamics of polymer fluids \cite{Lattice_1998_103_335, Lattice_2014_141_044909} to the AG model \cite{Temperature_1965_43_139} to obtain the GET \cite{Generalized_2008_137_125, Polymer_2021_54_3001, Thermodynamic_2023_41_1329, Advances_2023_53_616} is that this theory provides a precise method for calculating $S_c^*$ for any polymer material to the extent that the polymer material can be described by the united-atom lattice model of polymer melts. \cite{Generalized_2008_137_125} It has taken some time for us, however, to appreciate the physical significance of this basic fluid property because of its rather abstract nature. Numerical estimates of $S_c$ as a function of $T$ based on MD simulations \cite{Relationship_2013_138_12A541} in which minima of the energy landscape are sampled by inherent structure minimization and complementary explicit calculations of $S_c$ based on the GET have both provided some intuitive physical understanding of $S_c^*$, in addition to its quantitative estimation for model fluids.

Simulation estimates of $S_c$ under constant density conditions indicate that this quantity, along with the average potential energy calculated from inherent structure quenching, tends to generically approach a constant, or at least becomes slowly varying, at elevated temperatures. \cite{Relationship_2013_138_12A541} From an energy landscape perspective, we may understand $S_c$ as a measure of the total number of distinct thermodynamic states accessible to the fluid. This number is naturally finite in the lattice model because of the assumed lattice structure, and in the off-lattice case, the finiteness of $S_c$ owes its existence to the ultimately discrete nature of matter at molecular scales. Above some characteristic onset temperature $T_A$, \cite{Signatures_1998_393_554, relationship_2001_409_164} both $S_c$ and the inherent structure energy tend to saturate and this thermodynamic condition happens to coincide with the onset of non-Arrhenius dynamics at temperatures lower than $T_A$, providing an unequivocal method for estimating $T_A$. Under these thermodynamic conditions above $T_A$, the material system becomes insensitive to the highly ``corrugated'' nature of the energy landscape and the dynamics becomes strongly chaotic so that simple stochastic models of the material dynamics reasonably apply, such as Langevin models and classical TST. One might then say that the material is ``dynamically homogeneous'' in this restricted sense. Under constant pressure conditions, one has to consider a free energy surface rather than a potential energy surface in the minimization procedure, but one can formally define $S_c$ in a similar manner. Roughly speaking, $S_c$ is a counting function for the accessible minima of the energy surface at any given temperature, a basic metrical parameter of the energy landscape having simultaneously a thermodynamic and dynamic significance. Correspondingly, $S_c^*$ describes the complete tally of such inherent structure states, and thus describes the overall ``complexity'' or ``thermodynamic depth'' of the energy landscape. For further information about the energy landscapes and dynamics, we refer the reader to the highly readable paper by Debenedetti and Stillinger, \cite{Supercooled_2001_410_259} who also emphasize the theoretical challenge of estimating fragility of real materials. Regarding a compelling discussion of energy landscape from the standpoint of experimental scientists, we recommend the works of Ngai and Roland \cite{Chemical_1993_26_6824} and Santangelo and Roland, \cite{Molecular_1998_31_4581} who boil down this perspective to infer that fragility of polymers is directly related to their structure, i.e., smooth, compact, and symmetrical chains exhibit ``strong'' relaxation behavior, while fragile polymers are those with more rigid backbones or sterically hindering pendant groups, a view consistent with the GET predictions. Now that we have sketched the physical significance of $S_c^*$, we move on to explore its relation to the properties of model polymer fluids through its direct calculations based on the GET.

\begin{figure*}[htb]
	\centering
	\includegraphics[angle=0, width=0.9\textwidth]{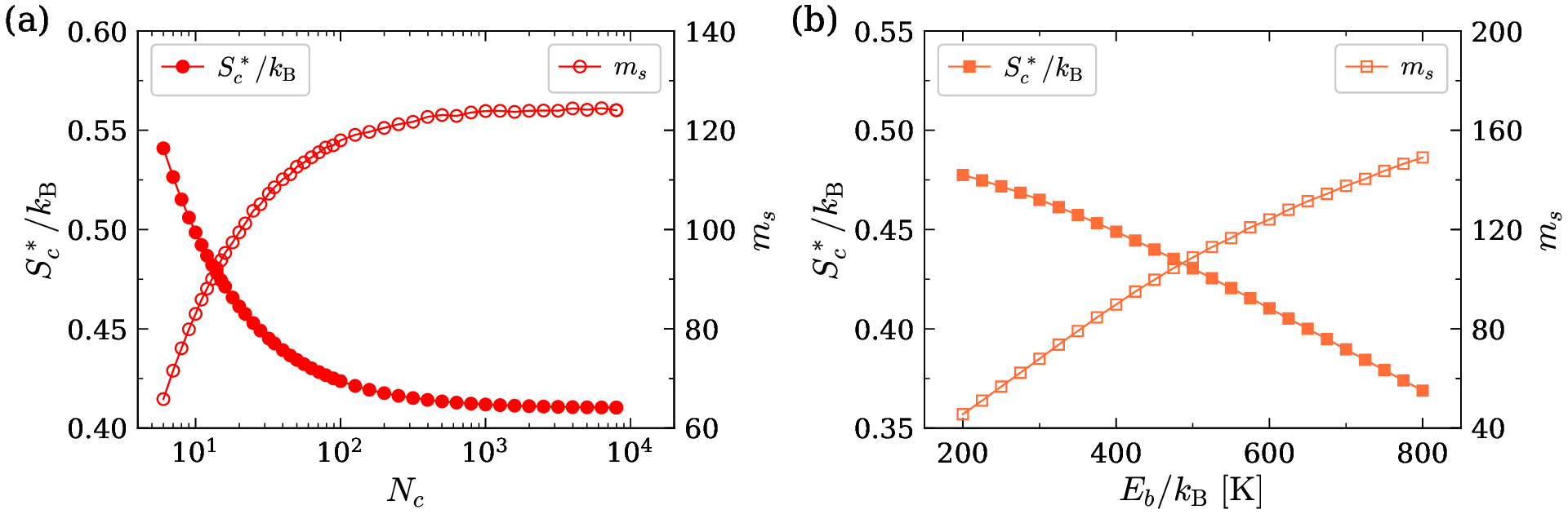}
	\caption{Molecular mass and chain stiffness dependences of the high temperature limit $S_c^*$ of the configurational entropy density and the segmental fragility $m_s$ predicted by the GET. Panels (a) and (b) show $S_c^*$ and $m_s$ as a function of $N_c$ and $E_b$, respectively. The calculations are performed for the PP structure.}\label{Fig_STA_PP1}
\end{figure*}

\begin{figure*}[htb]
	\centering
	\includegraphics[angle=0, width=0.8\textwidth]{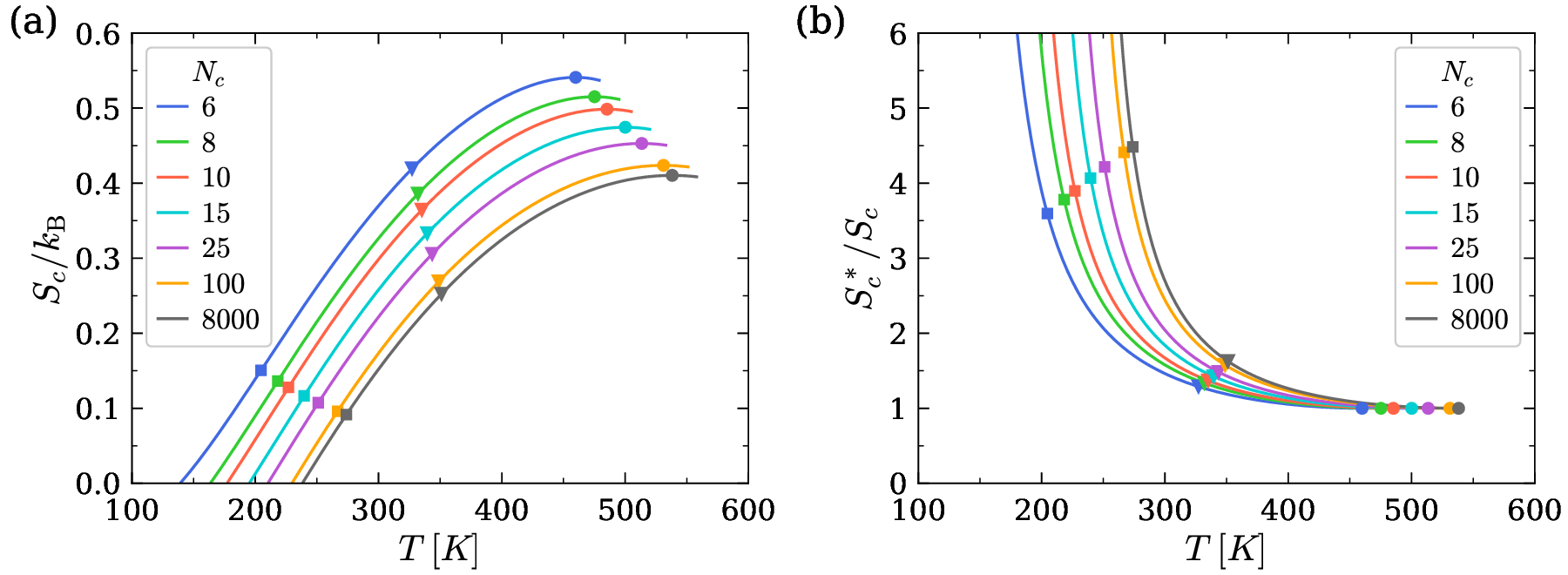}
	\caption{Temperature and molecular mass dependences of the configurational entropy density and the extent of cooperative motion predicted by the GET. Panels (a) and (b) show $S_c$ and $S_c^* / S_c$ as a function of $T$ for a range of $N_c$. Circles, triangles, and squares correspond to the positions of the onset temperature $T_A$, crossover temperature $T_c$, and glass transition temperature $T_{\mathrm{g}}$, respectively. The calculations are performed at variable chain length for the PP structure. We note that the vanishing of $S_c$ at low $T$ in the lattice mean field theory calculations is probably an artifact of the high temperature expansion of the polymer bending rigidity. This unphysical behavior does not occur when this approximation is avoided in the case of flexible polymer chains and instead $S_c$ is nearly constant at very low $T$, corresponding to a low temperature residual entropy. We think that this is the physically correct situation for semi-flexible polymers as well. \cite{Generalized_2016_145_234509, Polymer_2021_54_3001} Assuming that this low temperature limit for $S_c$ is general, the AG and GET models then both predict that Arrhenius relaxation should be recovered at very low $T$, at least under conditions where equilibration is possible.}\label{Fig_Sc}
\end{figure*}

We first focus on the simple PP structure that we have extensively studied in the past as a simple ``model'' polymer. \cite{Influence_2014_47_6990, Equation_2021_54_3247, Thermodynamic_2022_55_8699} Our direct computation indicates that $S_c^*$ decreases with increasing the molecular mass (Figure \ref{Fig_STA_PP1}a), so we may anticipate a significant increase in $m_s$ relative to the monomer form of this polymer, $m_o$. In consistency with this expectation, an examination of $m_s(N_c) / m_s(6)$ for this model yields a significant increase of $m_s$ arising from the polymeric nature of the molecule (Figure \ref{Fig_STA_PP1}a). Here, we select the chain with $N_c = 6$ as the model ``oligomer'', \cite{Role_1994_238_41} a polymer of minimal length. Indeed, we see that $m_s$ of a high mass polymer increases relative to its monomer value by a factor of about $2$. This finding confirms that the fragility of polymers is normally much larger than small molecule liquids. \cite{Why_2016_145_154901} Specifically, it has been stated as a general trend that the segmental fragility of polymers is about $1.5$ times the fragility of the most fragile small molecule GF liquid. \cite{Why_2016_145_154901} The GET predictions then appear to rationalize this observed trend in small molecule versus polymeric liquids well. We may appreciate the enhanced variation of fragility with increasing chain length from the corresponding plots of $S_c$ and $S_c^* / S_c$ as a function of $T$, as shown in Figure \ref{Fig_Sc}. It is apparent that as $N_c$ is increased, the strength of the $T$ variation of $S_c^* / S_c$ increases progressively with \textit{decreasing} $S_c^*$. The GET also provides insight into the variation of the fragility of polymers with chain stiffness. In Figure \ref{Fig_STA_PP1}b, we show the variation of $S_c^*$ with the chain stiffness parameter $E_b$, where we restrict to long polymer chains. Increasing $E_b$ also leads to a progressive decrease in $S_c^*$. Correspondingly, we see that $m_s$ increases in a monotonic way with increasing $E_b$ (Figure \ref{Fig_STA_PP1}b).

The decrease of the configurataional entropy density $S_c^*$ in the athermal limit with increasing polymer mass $N_c$ shown in Figure \ref{Fig_STA_PP1}a was not originally anticipated, so we now attempt to rationalize this trend revealed by the GET. In particular, we heuristically attribute the progressive decrease in $S_c^*$ calculated from the GET with increasing mass to correspond to a change of the configurational entropy per link of the polymer chains from a value similar to random walks when the chains are short to Hamilton walks (space filling self-avoiding polymers when the chains are long), this change in the entropy per link being well-approximated by $1/e$ in a mean-field approximation. \cite{From_1988_53_1139, Thermodynamics_1992_67_395, Self-avoiding-walk_1995_51_1791, Spectrum_1997_55_738} In the extreme of low molecular mass polymers and molecular liquids, we can qualitatively understand their relatively high configurational entropy relative to high molecular mass polymers from the configurataional entropy of space filling dimers. \cite{Exact_2008_100_120602} Butera et al. \cite{Higher-order_2013_87_062113} have estimated the dimer configurational entropy in $d$ spatial dimensions and Xu et al. \cite{Entropy_2016_161_443} have correspondingly estimated $S_c^*$ for high molecular mass polymer liquids in $d$ dimensions. These idealized models of ``athermal'' polymer and molecular fluids described by Hamilton walks and dimers are mentioned to give only a qualitative sense of why $S_c^*$ naturally decreases with increasing molecular mass. Of course, the relatively low configurational entropy of high molecular mass polymers is a well-known general attribute of these fluids that accounts for their often observed relatively low miscibility with other liquids having a similar chemistry.

\begin{figure*}[htb]
	\centering
	\includegraphics[angle=0, width=0.975\textwidth]{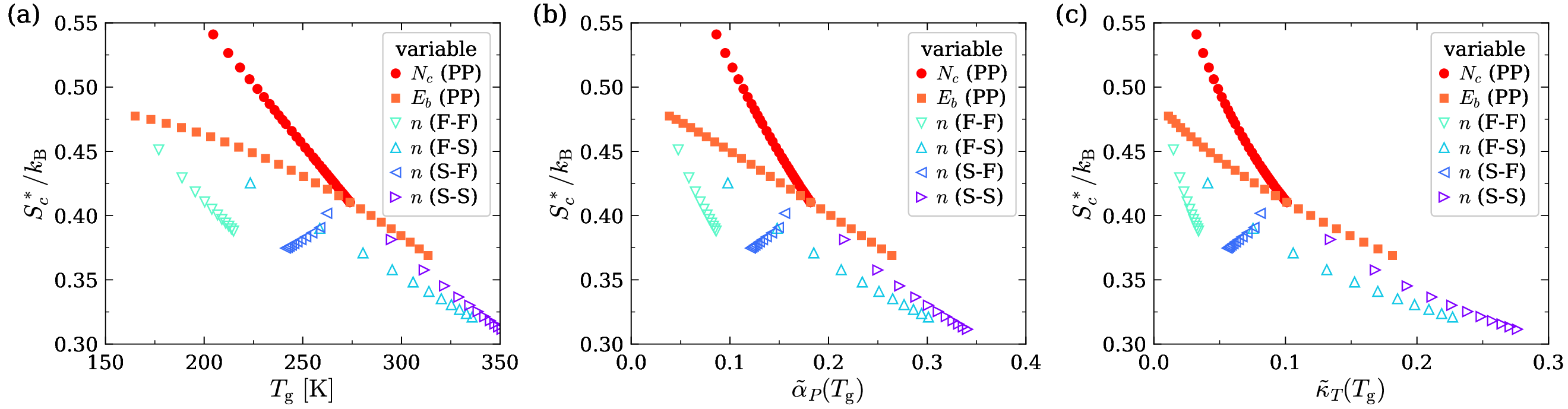}
	\caption{Correlation between $S_c^*$ and the glass transition temperature $T_{\mathrm{g}}$, the dimensionless thermal expansion coefficient $\tilde{\alpha}_P(T_{\mathrm{g}})$, and the dimensionless isothermal compressibility $\tilde{\kappa}_T(T_{\mathrm{g}})$ predicted by the GET for the PP structure having variable $N_c$ and $E_b$ and for different classes of polymers having variable $n$.}\label{Fig_STA_Merge1}
\end{figure*}

When we compare $S_c^*$ to $T_{\mathrm{g}}$ in Figure \ref{Fig_STA_Merge1}a, we see that there is a general trend for $S_c^*$ to decrease with all these quantities when $\epsilon$ and $P$ are fixed. A smaller $S_c^*$ tends to make the variation of $S_c$ with $T$ ``sharper''. \cite{Does_2006_125_144907} We may gain insight into the variation of these characteristic temperatures from the variation of $S_c^*$ with more accessible thermodynamic properties that reflect the structure of the energy landscape. Previous works have established that packing frustration plays a central role in influencing the fragility of glass formation. \cite{Generalized_2008_137_125} In particular, the dimensionless thermal expansion coefficient and isothermal compressibility
\begin{equation}
	\label{Eq_ThermoReduce}
	\tilde{\alpha}_P = T\alpha_P,\ \tilde{\kappa}_T = \rho k_{\mathrm{B}} T \kappa_T
\end{equation}
provide accessible experimental measures of packing frustration, \cite{Polymer_2021_54_3001, Entropy_2016_161_443} where $\rho = \phi / V_{\mathrm{cell}}$ with $\phi$ and $V_{\mathrm{cell}}$ being the filling fraction and the volume of a single cell, and the thermal expansion coefficient $\alpha_P$ and isothermal compressibility $\kappa_T$ follow the standard definitions,
\begin{equation}
	\label{Eq_Thermo}
	\alpha_P = \frac{1}{V}\left. \frac{\partial V}{\partial T} \right|_P,\ \kappa_T = -\frac{1}{V} \left. \frac{\partial V}{\partial P} \right|_T
\end{equation}

We see in Figures \ref{Fig_STA_Merge1}b and \ref{Fig_STA_Merge1}c that $S_c^*$ progressively decreases with increasing $\tilde{\alpha}_P (T_{\mathrm{g}})$ and $\tilde{\kappa}_T (T_{\mathrm{g}})$ at fixed $\epsilon$ and $P$ so that increased packing frustration as measured by these thermodynamic properties corresponds to a decrease of $S_c^*$, i.e., the energy landscape becomes more sparse and the glass formation becomes ``sharper'', pushing the characteristic temperatures to higher values. (The trend is different when $P$ and $\epsilon$ are varied over a large range, as we discuss in Section \ref{Sec_Inverted}.) The data for the S-F model with increasing the side-group length $n$ is an exception to this general trend. In this model, the extension of the flexible side groups relieves packing frustration (quantified below) and $S_c^*$ of a polymer melt of relatively stiff chains increases with increasing $n$.

Packing efficiency can be enhanced by other factors, such as increasing the external pressure or the intermolecular interaction strength, which increases the internal pressure of the polymer material, a variation that leads to similar effects on the polymer melt dynamics. We discuss this phenomenon in Section \ref{Sec_Inverted}, where different fragility changes are anticipated by changes in $S_c^*$.

A previous work based on the GET in $d$ spatial dimensions \cite{Entropy_2016_161_443} has established that the relation between $m_s$ and the thermodynamic frustration parameters $\tilde{\alpha}_P$ and $\tilde{\kappa}_T$ is quite general in polymer GF liquids. A near inverse scaling of the relation between $m_s$ and cohesive interaction strength is predicted by a more recent theory of glass formation by Zaccone and coworkers \cite{Interatomic_2015_112_13762, Disentangling_2017_95_104203} developed with metallic GF fluids mainly in view. The near universal scaling relationship between $m_s$ and $\tilde{\alpha}_P$ has recently been reported to hold for a remarkably large range of non-polymeric fluids. \cite{Thermal_2023_19_694} It would be interesting to better understand whether or not there is any direct relationship between the model of Zaccone and coworkers \cite{Interatomic_2015_112_13762, Disentangling_2017_95_104203} and the GET.

Finally, we mention as a practical matter that some polymer materials are engineered to extremely high packing frustration to serve as materials for membranes for the separation of molecules, including water filtration, hydrogen storage, and heterogeneous catalysis applications. These ``polymers of intrinsic microporosity'' \cite{Polymers_2006_35_675, Polymers_2006_245_403, Polymers_2012_2012_513986} often have $T_{\mathrm{g}}$ and fragility values that are so high that these materials are difficult to measure or defy measurement. \cite{Effect_2008_107_1039}

\subsection{\label{Sec_Inverted}Inverted Fragility Trends Arising from Varying the Cohesive Energy Density}

In Section \ref{Sec_STA}, we are concerned with an understanding of general trends in the segmental fragility, the extent of cooperative motion, and $T_{\mathrm{g}}$ in polymeric GF liquids under the ``normal'' conditions where the intermolecular interaction strength (described by $\epsilon$ in the GET model) is relatively constant (e.g., synthetic polymers with van der Waals interactions between the monomer segments) and constant external pressure $P$. The alteration of $P$ or the internal pressure through the alteration of the cohesive energy density can also influence molecular packing efficiency, which naturally leads to changes in the dynamics of GF liquids in a way quite different from what we have seen in Section \ref{Sec_STA}. Nonetheless, we can still understand general trends in the segmental fragility, the extent of cooperative motion, and $T_{\mathrm{g}}$ to arise from the impact of these perturbations on packing frustration, as quantified by $S_c^*$. Apart from understanding the obviously important problem of the effect of pressure and other applied stresses such as uniaxial compression in a lubrication setting to applied electric fields in the performance of polymer materials in electronic applications, this situation can also arise in the case of nanoparticle and solvent additives in polymer materials, e.g., the addition of strongly associated side groups to the polymer backbone as part of an effort to manufacture more recyclable materials, the confinement of the polymer material to nanoparticle, nanofiber, and thin film geometries, and even from altering the polymer topology from being linear to many-arm star, \cite{Confinement_2024_160_044503} knotted ring, \cite{Glass_2024_57_6875} and comb \cite{Melt_2022_55_3221} molecular structures. Basically, we can expect essentially any factor that impacts the cohesive energy density of the polymer material to act in a similar way, and we illustrate the general trend in this situation through specific examples. We believe that an understanding of these trends should be highly beneficial for engineering the properties of polymer materials for their intended applications.

\begin{figure*}[htb]
	\centering
	\includegraphics[angle=0, width=0.9\textwidth]{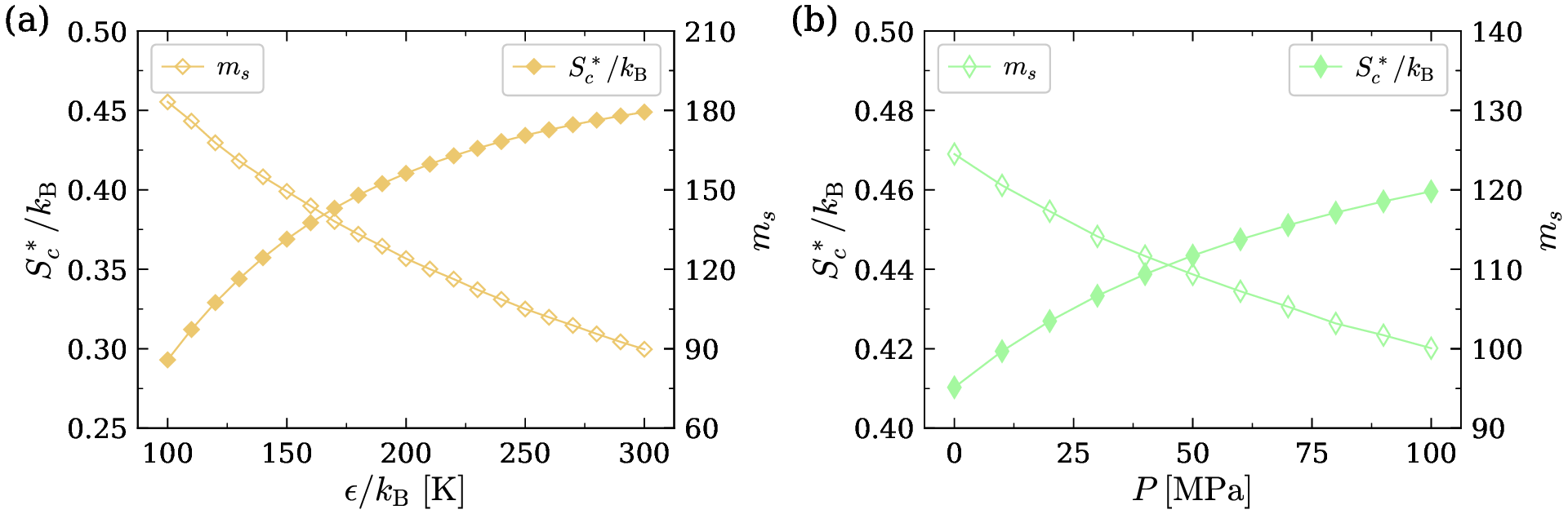}
	\caption{Cohesive interaction strength and pressure dependences of the high temperature limit $S_c^*$ of the configurational entropy density and the segmental fragility $m_s$ predicted by the GET. Panels (a) and (b) show $S_c^*$ and $m_s$ as a function of $\epsilon$ and $P$, respectively. The calculations are performed for polymer melts having the PP structure.}\label{Fig_STA_PP2}
\end{figure*}

\begin{figure*}[htb]
	\centering
	\includegraphics[angle=0, width=0.975\textwidth]{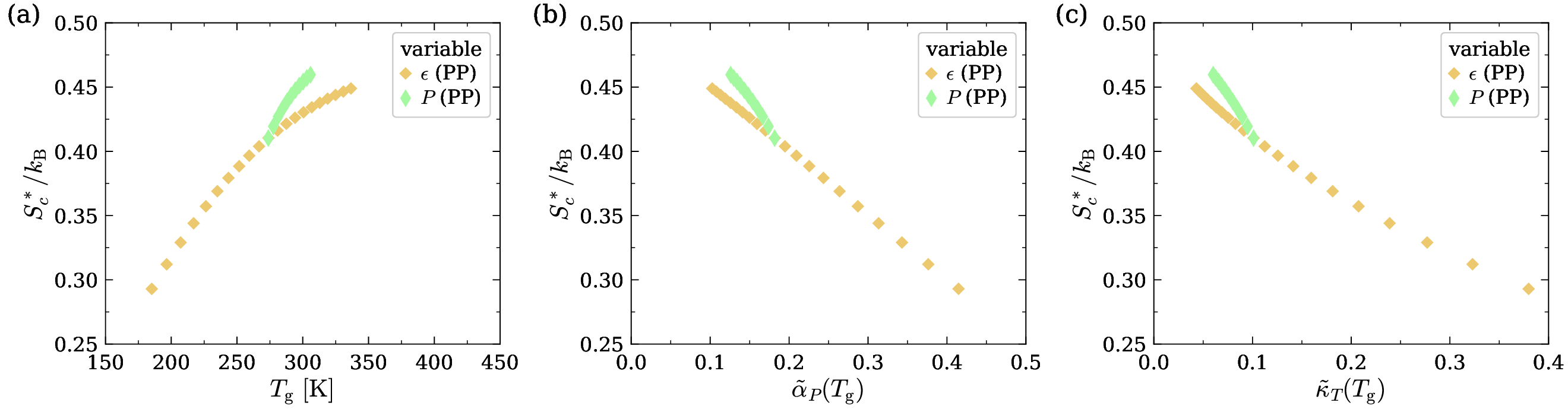}
	\caption{Correlation between $S_c^*$ and the glass transition temperature $T_{\mathrm{g}}$, the dimensionless thermal expansion coefficient $\tilde{\alpha}_P(T_{\mathrm{g}})$, and the dimensionless isothermal compressibility $\tilde{\kappa}_T(T_{\mathrm{g}})$ predicted by the GET for the PP structure having variable $\epsilon$ and $P$.}\label{Fig_STA_Merge2}
\end{figure*}

Varying the cohesive interaction parameter $\epsilon$, and thus the relative magnitude of the material cohesive energy density and internal pressure, \cite{Influence_2016_49_8341} as well as the applied external pressure $P$ have the \textit{opposite} effect of causing $S_c^*$ to progressively increase rather than decrease, as illustrated for our representative polymer models in Figure \ref{Fig_STA_PP2}a. Correspondingly, we show in Figure \ref{Fig_STA_PP2}b that the increase of these pressure related properties causes $m$ to decrease, as expected from the variation of $S_c^*$ and the discussion above.

Increasing $P$ and $\epsilon$ normally increases $T_{\mathrm{g}}$. If the changes in these variables are large, then they can lead to a decrease of $m_s$, a trend that is increasingly observed in functional polymers engineered for various applications. On the other hand, an increase in $P$ and $\epsilon$ leads to an increase of $S_c^*$, reflecting the enhancement of molecular packing efficiency, which naturally explains the reduction of $m_s$ from the GET perspective. We note that the initially unanticipated inverted trend between $T_{\mathrm{g}}$ and $m_s$ in the GET was first found by Stukalin et al. \cite{Application_2009_131_114905}

Many experimental studies for systems in which the cohesive interaction strength increases through the introduction of charged and polymer groups have correspondingly indicated a tendency of $T_{\mathrm{g}}$ to increase while fragility decreases as the molecular parameter influencing the cohesive interaction strength of the fluid increases. These systems include melts of polyzwitterion materials having increasing charge density, \cite{GlassForming_2021_54_10126} polymer networks with strong hydrogen bonding, \cite{Tuning_2015_48_4196, Hydrogen_2020_201_122627, Tuning_2021_17_7541, Heterogeneous_2023_56_4336} weakly coordinating ionic physical cross-links, \cite{Tuning_2022_55_9478} and ionomers of increasing charge density, \cite{Unusual_2022_55_6536} where the decreased fragility is implied qualitatively by a significant broadening of the breadth of the glass formation. Less obviously, the variations of $T_{\mathrm{g}}$ and fragility in highly knotted ring \cite{Glass_2024_57_6875} and many-arm star \cite{Confinement_2024_160_044503} polymer melts appear to follow this general pattern of behavior, a result attributed to the effect of high topological complexity to the internal pressure of the polymer fluid. \cite{Glass_2024_57_6875} Xia and coworkers \cite{Competing_2022_55_9990} have recently shown that $T_{\mathrm{g}}$ increases while fragility decreases in MD simulations of semiflexible polymer networks with increasing $\epsilon$.

The inverted trend between the variations of $T_{\mathrm{g}}$ and $m_s$ is a characteristic feature of the ``antiplasticization'' phenomenon in polymers, which results from adding molecular additives of high cohesive energy, among other properties related to molecular geometry, to polymer melts. \cite{Influence_2006_97_045502, Tuning_2007_126_234903, Antiplasticization_2010_6_292, additive_1984_29_1403, Antiplasticization_2020_12_769} Interestingly, the addition of some types of nanoparticles (e.g., fullerene particles) to polymer materials [e.g., polystyrene (PS)] can lead to an increase of $m_s$ and a decrease of $T_{\mathrm{g}}$ at low concentrations of the nanoparticle additive where significant nanoparticle aggregation is avoided. \cite{Glass_2010_153_79, Influence_2015_68_47} Formally, this effect might be called ``anti-antiplasticization'' for the lack of an established term for this singular phenomenon in which the additive disrupts molecular packing and increases rather than decreases the fragility of glass formation, as in antiplasticizing additives.

There are an extraordinary number of applications that exploit this phenomenon, e.g., the addition of additives to make polymer films scratch resistant, \cite{Antiplasticization_1992_25_4588} the enhancement of the shelf life of protein drugs, \cite{Relaxation_2012_8_2983, Dielectric_2006_74_031501, Quantifying_2008_112_15980} the stabilization of lithographically etched polymer nanostructures, \cite{Mechanical_2010_6_2475} and the mechanical properties of paper and other diverse organic materials including foods. \cite{Antiplasticization_1999_64_576, Mathematical_1996_73_712, Mapping_1993_32_575, Mechanical_2004_93_2896, Plasticization_2018_21_72, Antiplasticization_2008_106_1417, Antiplasticization_2006_73_1, Plasticizing_2000_65_445, Solvent_2013_9_5336} This overall effect is simple to understand from the GET. The small molecule additive (or judiciously chosen polymer side groups) enhances the cohesive energy density of the polymer and molecular packing so that segmental fragility drops. The corresponding increase in the cohesive energy density, on the other hand, makes $T_{\mathrm{g}}$ rise. In ordinary ``plasticization'', the additive causes a general tendency that both $T_{\mathrm{g}}$ and $m_s$ decrease in tandem. It is worth noting that the incorporation of charges, molecules, or subunits into the material is an effective way to enhance the cohesive energy density \cite{Influence_2021_54_9587, Molecular_2023_56_4049} for applications of the type just mentioned.

We have seen that the inverted trend between $T_{\mathrm{g}}$ and $m_s$ can arise in more subtle contexts in which the polymer topology is altered. In particular, we have observed this type of inverted trend in simulated many-arm star \cite{Confinement_2024_160_044503} and knotted ring \cite{Glass_2024_57_6875} polymers when the topological complexity becomes large and gives rise to a large change in the internal pressure of the fluid and a significant increase in the average density of the polymer material arising from these topological constraints. These changes in $T_{\mathrm{g}}$ and $m_s$ are directly supported by corresponding predicted changes in the dimensionless thermal expansion coefficient and isothermal compressibility. The inverted trend has been observed experimentally upon introducing ionic physical cross-links to a polymer material. \cite{Heterogeneous_2023_56_4336} We can expect many further applications exploiting this effect to engineer polymer properties in the future.

We finally note that while almost all synthetic polymers apparently follow the fragility trend described in Section \ref{Sec_STA}, unless molecular, nanoparticle, or strong physical cross-links intervene in altering this pattern of behavior, there is an interesting, commonly encountered, and practically important polymer that appears to follow the inverted trend between $m_s$ and $T_{\mathrm{g}}$ described above---polyisobutylene (PIB). \cite{Polyisobutylene_2008_46_1390} Even though PIB would seem to be from a chemistry standpoint just a ``garden variety'' polyolefin polymer having a remarkably simple monomer structure, this polymer exhibits highly exceptional properties that have led to manifold applications exploiting them, such as low permeability to air and small molecules, as a synthetic rubbery material (isobutyl rubber) of profound practical importance, and as a lubricant, and in biomedical applications. \cite{Atomistic_2008_41_6228, Anomalous_1995_28_1252} From the standpoint of the present paper, this polymer is notable for having one of the lowest fragilities of all polymer materials. \cite{Atomistic_2008_41_6228, Nonexponential_1993_99_4201} While the precise origin of the special properties of PIB is another enduring mystery of polymer science, Kunal et al. \cite{Polyisobutylene_2008_46_1390} have thoughtfully reviewed many of the special molecular properties of this polymer possibly contributing to its special dynamical properties, and Ding et al. \cite{Influence_2004_37_9264} have discussed many other ``special'' properties of PIB in comparison with other polymers. In particular, Kunal et al. \cite{Polyisobutylene_2008_46_1390} emphasized that the presence of methyl groups attached to every other carbon atom of the chain backbone causes the \textit{trans} and \textit{gauche} configurations of the polymer to have nearly the same potential energy, an effect that is responsible for the coiled nature of the polymer in comparison to polyethylene. Apart from explaining the high elasticity of PIB, this molecular coiling phenomenon has also been often suggested to account for the low propensity for this polymer to crystallize, as most other polyolefins readily do. It was further observed by Kunal et al. \cite{Polyisobutylene_2008_46_1390} that this energetic degeneracy in polymer conformational states also leads to a polymer fluid having an exceptionally large configurational entropy compared to other polyolefins. This observation, following the discussion above, points to a material that should have an exceptionally low fragility, which is certainly the case for PIB. We also think that this is a key observation relevant for understanding the dynamical properties of PIB, and butyl rubber materials derived from this polymer. However, the demonstration of this hypothesis requires more detailed modeling based on a more atomistic treatment of PIB polymer structure to fully address the problem.

Here, we note some of the essential observed properties of PIB indicating that this polymer should indeed belong to the family of materials described in this section. PIB is also characterized as having the highest density of all polyolefins and exceptionally low thermal expansion coefficient and isothermal compressibility. \cite{Anomalous_1995_28_1252} Systematic studies of blends of PIB with other polyolefins further indicate that the polymer interaction parameter of PIB with other polyolefins is normally large in magnitude and negative,implying the presence of a highly attractive cohesive interaction suggested to arise from some sort of ``local packing contribution'', given that there does not appear to be any obvious source of a specific chemical interaction that might explain the high molecular cohesion in this material. \cite{Anomalous_1995_28_1252} Given this strong attractive cohesive interaction, regardless of its source, it is then no surprise from the standpoint of the GET that $T_{\mathrm{g}}$ increases, but $m_s$ decreases, with increasing polymer mass in PIB. \cite{Role_2010_43_8977, Polyisobutylene_2008_46_1390}The same trend with varying polymer mass has been observed previously in MD simulations of polymers having high topological complexity, such as many-arm star \cite{Confinement_2024_160_044503} and knotted ring \cite{Glass_2024_57_6875} polymers, where it was found that the strong topological complexity increases the fluid internal pressure and thus leads to an inverted variation between $T_{\mathrm{g}}$ and $m_s$. We suggest that this is exactly what is going on in PIB, although the exact nature of the topological interaction that might be responsible is frankly obscure. Could the polymer be exhibiting some propensity to form helical or other special organized structures in the melt? A full understanding of what physically makes PIB so special in its properties thus remains somewhat of a mystery. 

\subsection{\label{Sec_Length}Dependence of Fragility on Observational Length Scale in Polymer Liquids}

As mentioned in Section \ref{Sec_Intro}, another awkward aspect of the dynamics of polymers, and presumably many other complex fluids, is that the structural relaxation time can depend strongly on length scale. In particular, relaxation times associated with polymer chain diffusion and the shear viscosity of the entire fluid, a property defined in the thermodynamic limit, can exhibit a very different temperature dependence than the segmental relaxation time. Evidently, fragility $m_c$ at the scale of the polymer chains has experimentally been found to be relatively strong. \cite{Why_2007_19_205116, Role_2010_43_8977, Why_2016_145_154901, Surprising_2018_51_4874, New_2006_39_8867} Studying this phenomenon is complicated by the fact that it is difficult to determine large-scale chain diffusion and relaxation processes in polymer melts at low $T$ because of the extremely large computational times involved, resulting in a dearth of computational information about large-scale polymer chain dynamics under such conditions. This is another fundamental unresolved problem regarding the dynamics of GF polymer liquids. Here, we offer a tentative model about how this important phenomenon might be addressed within the GET theoretical framework.

While the apparent ``structure'' of a polymer melt depends on the scale of observation, the thermodynamic properties of the bulk polymer materials are invariant to the choice of length scale, so we make our arguments about the scale dependence of chain dynamics based on this simple observation. To make our arguments concrete, we start by considering a typical description of a polymer model treated ``realistically'' at a segmental scale. A polymer model involving a flexible polymer with stiff side groups (the so-called F-S model in the GET \cite{Generalized_2008_137_125, Polymer_2021_54_3001}) provides a representative example of such a model for the purpose of illustrating general ideas. (Alternatively, one could start with a fully atomistic description of the polymer in which every atom is considered, or a united-atom model in which some atoms, such as H atoms, are neglected, but such descriptions would be difficult to implement in the GET, which is admittedly a coarse-grained model. The detailed description of any particular polymer is not our concern in the present work, since our aim is to understand general property trends in polymers.) The realization of the F-S model polymer is epitomized by the case of PS, which is a ``typical'' polymer GF liquid having a segmental dynamics that is relatively fragile. Now, if we view the same polymer at a larger length scale comparable to the scale of the polymer chains, then it seems reasonable to describe the \textit{same} polymer as a string of ``blobs'' containing many monomer units. At this scale, the polymer is highly flexible, which is a defining characteristic of the Rouse theory, so a chain of such coarse-grained segments can reasonably be described as a random walk of such segments. This discussion necessitates chains long enough for many such coarse-grained units to exist. Formally, we may equally as well describe the PS chains within the GET as a simple chain of coarse-grained spherical beads without side groups, similar in spirit to the work of Guenza and coworkers, \cite{Coarse-graining_2018_14_7126} and having the same polymer-polymer interaction strength ($\epsilon$), as in the atomistic F-S model description of PS on a segmental scale. The size of the beads of the \textit{observationally coarse-grained polymer} must evidently be larger than the monomers in the segmentally defined chain, so the ``dynamical persistence length'' having a complex monomer structure can be expected to be correspondingly larger than the persistence length defined on the atomistic scale of the monomers. \cite{Size_2010_43_9126} The molecular parameter $E_{b,c}$ in the GET can be specified by demanding that $S_c$ extrapolates to $0$ at the same $T$, as in the polymer model described at the segmental scale. This thermodynamic invariance condition leads to $E_{b,c} / k_{\mathrm{B}} = 385$ K for the F-S polymer having $E_b / k_{\mathrm{B}} = 350$ K and $E_s / k_{\mathrm{B}} = 800$ K, as discussed in more detail below. We physically interpret this condition in our model as corresponding to the occurrence of an amorphous solidification transition in the material in which the material becomes effectively hyperuniform. \cite{Influence_2016_49_8341, Influence_2017_50_2585, Confinement_2024_160_044503, Glass_2024_57_6875} An important consequence of this working model of relaxation on a larger length scale is that both the monomer segmental and dynamical chain segmental relaxation times share the same VFT functional form with the same VFT temperature $T_0$, where the fragility parameters $K_c$ and $K_s$ of the segmental and chain relaxation processes are generally different.

Now, if we assume that the GET also applies to the polymer dynamics at a mesoscale where the polymer is described as a simple string of Gaussian beads, then we can estimate the relaxation time on this larger length scale relevant to the dynamics of the chain normal modes, large scale chain diffusion, and the shear viscosity of the polymer liquid. Before directly estimating this dynamical segmental relaxation time using the GET, we emphasize that this time is independent of polymer mass for chains long enough to be described by the Rouse and Reptation models. Thus, this dynamical segmental relaxation time cannot be directly identified with the strongly mass dependent long-time structural relaxation time of the polymer melt or the diffusion coefficient of the polymers chains. Rather, we identify this time with the often observed, but poorly understood ``sub-Rouse modes'' in fragile polymer GF liquids, which are likewise observed to be independent of polymer molecular mass and insensitive to polymer microstructure. \cite{Origin_2014_10_9324} Our estimation of the fragility of the ``dynamical segmental relaxation time'' or simply the ``chain segmental relaxation time'' $\tau_c$ is relatively low, compared to the ``monomer segmental relaxation'' or ``segmental relaxation time'' $\tau_{\alpha}$ in fragile GF polymer fluids. It is anticipated that that $\tau_c$ governs the $T$ dependence of the friction coefficient of the even more coarse-grained Rouse and Reptation models, but this relationship remains to be established quantitatively. 

\begin{figure*}[htb]
	\centering
	\includegraphics[angle=0, width=0.45\textwidth]{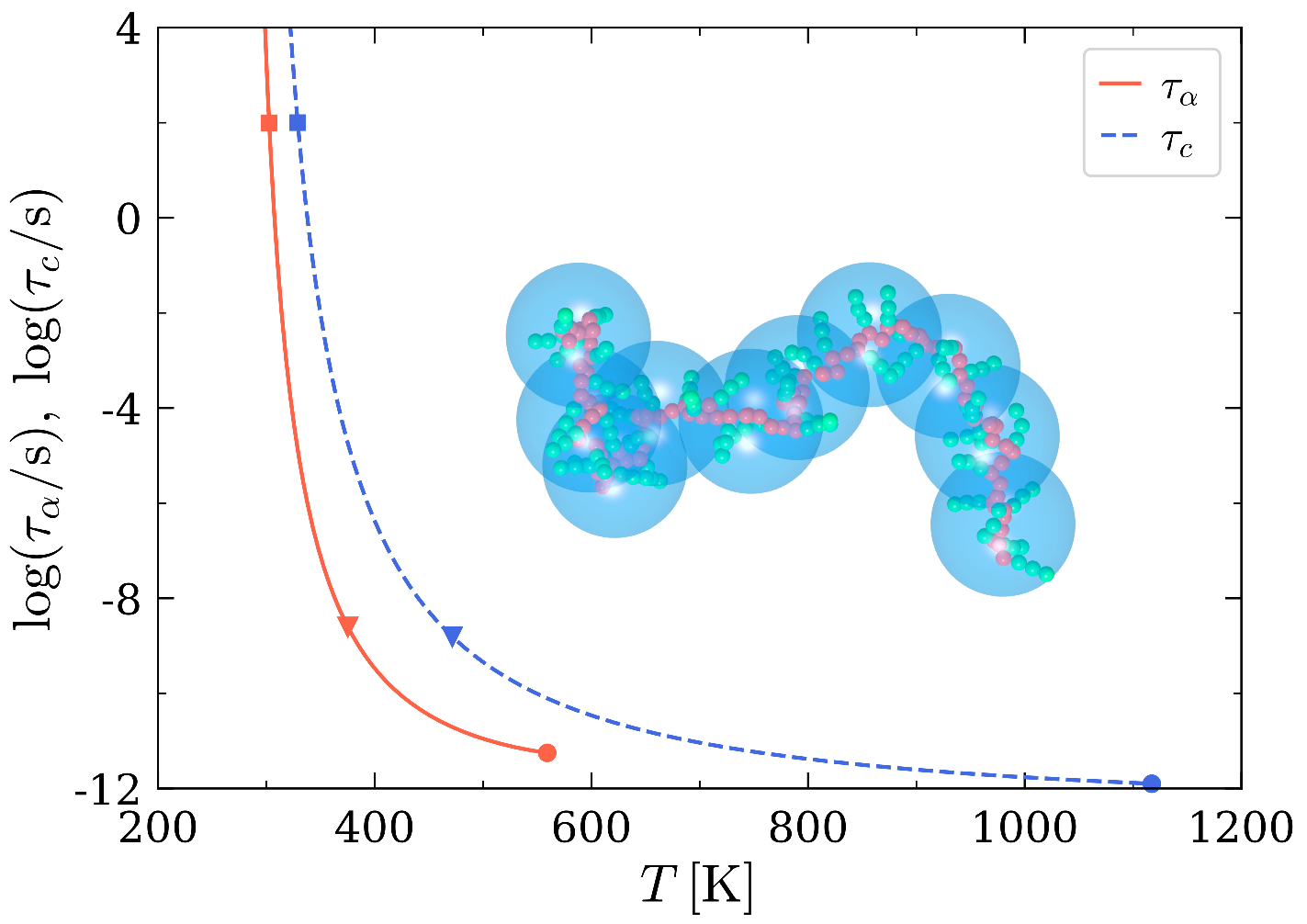}
	\caption{Monomer segmental relaxation time $\tau_{\alpha}$ and chain segmental relaxation time $\tau_c$ as a function of $T$ predicted the GET for the flexible-stiff coarse-grained polymer melt. Circles, triangles, and squares correspond to the positions of $T_A$, $T_c$, and $T_{\mathrm{g}}$, respectively. The inset shows the cartoon of a representative configuration of a polymer described by flexible backbone and relatively stiff side groups with a length of $n = 3$.}\label{Fig_TauCG}
\end{figure*}

We next consider an explicit calculation of $\tau_c$ based on the argument just sketched. As a representative model of the polymer chain, we adopt a polymer model from the class of polymer chains having a flexible backbone and stiff side groups, which has been taken as a model of PS and other relatively fragile polymers where this broad physical classification applies. Specifically, the chain backbone is taken to have a bending energy $E_b / k_{\mathrm{B}} = 350$ K, the side groups are assumed to have a modest size of three units ($n = 3$) with relatively stiff bonds ($E_s / k_{\mathrm{B}} = 800$ K) to roughly model the phenyl groups of PS, the cohesive interaction parameter is taken to be $\epsilon / k_{\mathrm{B}} = 200$ K, and the mass is taken to be relatively high, $N_c = 10000$, so that $T_{\mathrm{g}}$ and fragility safely have no mass dependence. A representative configuration of such a chain is shown in the inset of Figure \ref{Fig_TauCG}, where the structure at a monomer scale is indicated. As sketched previously, we then consider the same polymer chain on a lower scale of observation, where the chain is viewed as being comprised of ``blob-like'' structural units that we describe as simple ``beads'' in the GET. A similar coarse-graining idea underlies the Rouse and reptation models, \cite{Viscoelastic_1980_} where we also encounter the same question of just how many monomers of the polymer chain described at an atomistic scale belong to the ``beads'' of the coarse-grained polymer chain. Sokolov and coworkers \cite{When_2004_37_161, Size_2010_43_9126} have discussed this matter extensively where they found that the ``dynamical segments'' can be composed of as many as $500$ to $5000$ atomic units of the polymer, corresponding to chain sub-units much larger than that defined by the statistical segment length traditionally invoked to describe the static configurational structure of the polymers. Tentatively, we choose what we believe to be a representative value of $400$ atomic units so that our observationally coarse-grained polymer is then modeled at larger scales as a linear polymer composed of $25$ spherical beads. The beads describing the polymer in its coarse-grained description are evidently larger than the monomers in the segmentally defined chain. As described briefly above, $\tau_c$ can then be specified in the GET model by demanding that $S_c$ in the coarse-grained polymer model extrapolates to $0$ at the same $T$ as in the polymer model specified on a monomer scale, as discussed above. Within the GET, this condition leads to the precise value of the bending energy of the coarse-grained chain, $E_{b,c} / k_{\mathrm{B}} = 385$ K. The practical consequence of this working hypothesis is that the segmental and chain relaxation times ($\tau_{\alpha}$ and $\tau_c$) share the same VFT temperature, i.e., solidification does not depend on the length scale of observation. It is stressed that the GET predicts that both the dynamics on a monomer scale and on the larger scale of the coarse-grained dynamical segments exhibit a VFT functional dependence in the low temperature regime of glass formation below $T_c$ of the monomer segmental dynamics.

Figure \ref{Fig_TauCG} shows the GET predictions for $\tau_{\alpha}$ and $\tau_c$ implied by these arguments. The prefactor in the GET in this expression is still assumed to equal $10^{-13}$ s, although this constant might change in the coarse-grained model. The prefactor can be estimated from the calculation of the velocity autocorrelation function in the polymer model with and without coarse-graining. \cite{Chemically_2024_4_1018} This detail is ignored in the present work. Moreover, as noted above, the relation of $\tau_c$ to the characteristic normal mode times of the Rouse or reptation model is admittedly unclear. This complication is also neglected in our tentative model of chain segmental dynamics, where our interest is in qualitative trends. The GET and the stated assumptions predict that both the chain and segmental relaxation times should follow the VFT functional form in the low $T$ regime of glass formation, the regime of greatest practical interest for polymer materials, and the fact that the VFT temperature is the same for these relaxation times in our model, together imply that these relaxation times should be related by a power-law relation in the low temperature regime of glass formation below $T_c$, as observed experimentally in the low temperature regime of glass formation near $T_{\mathrm{g}}$. \cite{Effect_2012_45_8430}

Evidently, as shown in Figure \ref{Fig_TauCG}, the $T$ dependence of the relaxation time is much stronger on the scale of the monomers than on the scale of the polymer dynamical beads in our model fragile polymer. As noted before, we anticipate that this larger scale relaxation process should determine the temperature dependence of the chain normal mode dynamics, the polymer diffusion coefficient $D$ in the melt, and the shear viscosity $\eta$ of the polymer liquid through its relation to the chain friction coefficient in coarse-grained polymer models. Moreover, quantitative estimates of the ``segmental fragility'' and the ``chain fragility'' from $D$ and $\eta$ measurements are in reasonable accord with experimental observations. In particular, the estimated monomer segmental fragility of our representative fragile F-S polymer equals $m_s = 141$, while the chain segmental fragility equals $m_c = 85$, a relative fragility reduction of roughly $40 \%$. These estimates are compared with corresponding estimates for high molecular mass PS, where equal $m_s \approx 140$, while the chain fragility is about $m_c \approx 80$, \cite{Surprising_2018_51_4874} a relative fragility reduction having roughly the same relative magnitude.

We note that our model is completely in line with the arguments given before by Ding et al. \cite{Dielectric_2009_42_3201} to rationalize the change between the segmental and chain dynamics, as quoted here. ``According to this theory (GET), when there is a significant degree of packing frustration, the fragility of glass formation should be higher and the transport properties are more sensitive to temperature or other perturbations that can influence the packing geometry. The segmental motions are dominated by strong local hard core repulsive interactions that limit the efficiency of local packing. At longer time scales (on the order of $\tau_1$) and larger spatial scales (on the order of $R_{\mathrm{g}}$), the center of mass motion of the chains can be roughly described as interacting soft spheres because of their strong interpenetration. From this perspective, the dynamics of flexible chains at the scale of $R_{\mathrm{g}}$ should be universally that of a strong glass-former (showing weak temperature dependence of $\tau_1$) because of the relatively high packing efficiency of the soft spheres in comparison with hard spheres and other less penetrable particles. Moreover, the difference between the monomer segmental and chain relaxation processes should be more prevalent in systems whose segmental relaxation process is characterized by a more fragile glass formation, associated with a high local packing frustration due to strong local excluded volume interactions.''

There is another distinct aspect of the dynamics of unentangled polymer melts at large length scales because of the relative homogeneity of the fluid, the rate of molecular diffusion obtained from the rate of decay of the self-intermediate scattering function at long times and low wave numbers should scale inversely with the fluid viscosity. Consistent with this expectation, Urakawa et al. \cite{SelfDiffusion_2004_37_1558} have observed that the chain self-diffusion normalized by the thermal energy, $D_c / k_{\mathrm{B}} T$, of low molecular mass PS melts over a large $T$ range exhibits essentially the same temperature dependence as the reciprocal zero shear viscosity $1 / \eta$, confirming the absence of decoupling at larger length scales and the obeyance of the Stokes-Einstein relation for chain diffusion. Interestingly, a transition to the decoupling relation, $D_c \sim (1 / \eta)^{0.71}$, is found to occur when flexible polymers become long enough to become ``entangled'', \cite{Chain_2003_41_1589} strongly suggesting that dynamic heterogeneity reemerges in entangled polymer fluids even at chain length scales in association with entanglement. \cite{Weak_2018_4_19} Dynamic heterogeneity is also evidenced by the breakdown of the Stokes-Einstein relation of nanoparticle diffusion in polymer nanocompsites \cite{Breakdown_2007_7_1276} and a second peak in the non-Gaussian parameter on a time scale whose order of magnitude is related to the diffusion of the chain center of mass to a distance comparable to molecular dimensions. \cite{Localization_2014_89_052603} A similar second peak in the non-Gaussian parameter arises in simulations of polydisperse unentangled linear polymers, \cite{How_2022_55_9901} where the peak arises simply from the distribution of the mobilities of the polymer chains.

Overall, the tentative model discussed above suggests that the GET can be extended to describe chain segmental dynamics for the diverse applications in which polymer dynamics on large scales is required. Of course, some questions remain regarding how to precisely define the number of dynamic beads in the modeling of polymer chains, but this difficulty is not limited to modeling polymeric GF liquids. \cite{When_2004_37_161, Size_2010_43_9126}

\section{\label{Sec_Summary}Conclusion}

The high fragility of segmental relaxation in some polymer glass-forming fluids in comparison to molecular and atomic glass-forming liquids has been a long-standing puzzle that we directly address within the framework of the generalized entropy theory (GET) of polymer glass formation. This highly predictive theory of glass-forming liquids makes precise predictions on fragility of polymer fluids in terms of molecular structural parameters (e.g., chain length, rigidity, intermolecular interaction strength, the presence of side groups, etc.) and thermodynamic conditions such as variable pressure. The GET predicts that the monomer segmental fragility of a high mass fragile polymer glass-forming liquid, as defined by the steepness index $m_s$, can be a factor of $2$ or even larger than the corresponding value of a fluid composed of monomers having the same chemistry. This relatively high fragility in semi-flexible high molecular mass polymers at constant pressure and van der Waals interaction strength can be traced in the GET to the reduction of the configurational entropy density relative to the monomer fluid in the high temperature fluid state where $S_c$ can be taken to be constant $S_c^*$ to a reasonable approximation. A similar behavior arises in fluids exhibiting equilibrium polymerization upon cooling where the sharpness of the self-assembly transition, \cite{Lattice_2008_128_224901} defined in terms of the temperature range over which the thermodynamic transition occurs, increases as the ``entropy gap'' between the fluid configurational entropy in the high temperature ``dynamically homogeneous'' fluid state and the configurational entropy of the fluid in the fully associated low temperature state. \cite{Does_2006_125_144907} The fluid with a very high value of the configurational entropy in its high temperature dynamically homogeneous state can evidently accommodate a relatively large change in its configurational entropy before an entropy crisis intervenes that rigidifies the material. Materials having low values of configurational entropy tend to be more fragile, corresponding to an abrupt formation of dynamic heterogeneities and a corresponding abrupt change in the dynamics of the material. The suggestion here is that formation of dynamical structures in cooled glass-forming liquids corresponds to a kind of supramolecular assembly process. \cite{Does_2006_125_144907}

The usual pattern of behavior observed in polymers, e.g., by increasing the polymer chain length, molecular rigidity, or the complexity of the structure of the side groups so that the packing frustration of the molecules is increased and the configurational entropy in the athermal high temperature limit is reduced, and the segmental fragility and the glass transition temperature are both increased by the mechanism described above. It is exactly this situation that explains the Williams-Landel-Ferry (WLF) \cite{Temperature_1955_77_3701, Viscoelastic_1980_} ``universal scaling'' of the relaxation times. \cite{Meaning_2015_142_014905} In particular, a linear scaling between $T_{\mathrm{g}}$ and fragility underlies the empirical WLF relation with universal constants suitable for typical synthetic polymers.

The similar pattern of behavior no longer applies when either an applied pressure $P$ or the strength of the intermolecular interaction (quantified by $\epsilon$ in the GET model), which influences the fluid internal pressure, are varied over a large range. Increasing these variables causes $T_{\mathrm{g}}$ to increase, but the segmental fragility $m_s$ normally decreases. Again, this variation can be understood from how these variables influence the the configurational entropy in the athermal limit. There are a large number of systems that exhibit this inverted variation of between $T_{\mathrm{g}}$ and segmental fragility arising from any perturbation of the fluid that greatly enhances the cohesive energy density, while leaving other molecular parameters relatively fixed.

The definition and determination of fragility in polymeric, and presumably many other complex liquids, imply that this quantity depends on observational length scale. In particular, the fragility as defined on the ``segmental'' length scale of the monomers has often been observed to be quite different from the fragility at the scale of the large scale collective modes of the chains in polymers for which the segmental dynamics is fragile. This disparity in relaxation times at different length scales leads to a breakdown of the time-temperature superposition \cite{Temperature_1965_69_3480, Temperature_1982_20_729, Breakdown_2006_39_3322} and considerable uncertainty about how to model the polymer friction coefficient in familiar coarse-grained models of polymer melts, such as the Rouse and the reptation models. We introduce a tentative model to desribe the temperature dependence of the relaxation process on the length scale of the polymer chains based on the GET model. In this tentative model, we describe polymer chains as being composed of coarse-grained ``dynamical beads'' \cite{Size_2010_43_9126} when the polymer chain is viewed on larger length scales. This bead model description of polymers is consistent with the conventional qualitative physical picture underlying coarse-grained models of polymer dynamics, such as the Rouse and reptation models. Our model of chain segmental relaxation seems to reproduce many of the observed trends in fragility, as measured in large scale properties such as the polymer diffusion coefficient and shear viscosity, including the inherent tendency of the fragility to be relatively strong and insensitive to polymer chemistry and molecular mass on these larger length scales. Importantly, the predictions of the chain relaxation time $\tau_c$ is made within the same theoretical framework as the formerly developed GET model of the structural relaxation and thermodynamics of polymer fluids. Although the predictions of the model appear to be highly promising in the case of polystyrene, the new model of chain relaxation needs to be tested against measurements on other polymer materials to further check the hypotheses on which the model is based. The precise relation between the chain segmental relaxation time and normal mode relaxation times of the Rouse and repation models also needs to be better understood.

An unexpected finding of our work is that the segmental fragility $m_s$ can be directly related to the extent of collective motion near the glass transition temperature in the GET model, $S_c^* / S_c(T_{\mathrm{g}})$. Not only can the large fragility values of some polymers be readily rationalized by the GET, but we also discovered an unexpected near universal exponential relation between $m_s$ and $S_c^* / S_c(T_{\mathrm{g}})$, along with a minimal segmental fragility around $15$. While a literal interpretation of the diffential activation energy (local slope of the Arrhenius plot of the relaxation time) with the ``activation energy'' of the liquid would lead to a linear scaling of fragility with $S_c^* / S_c(T_{\mathrm{g}})$, the fact that the differential activation energy can be quite different from the actual activation energy makes this argument highly uncertain. Nonetheless, we find that $m_s$ is related to $S_c^* / S_c(T_{\mathrm{g}})$ to a good approximation, but just not in a linear way. Importantly, this scaling relation also explains why the segmental fragility is similar to atomic and molecular fluids in the limit when the polymers are oligomeric or composed of highly flexible polymer chains. Further, the general expression relating $m_s$ to $S_c^* / S_c(T_{\mathrm{g}})$ seems to be entirely consistent with known trends for polymer materials so that it is not necessary to introduce revised models of polymer glass formation to describe polymer materials, as has been suggested previously. \cite{Why_2007_19_205116, Role_2010_43_8977, Why_2016_145_154901, Are_2015_27_103101, Breakdown_2006_39_3322, Temperature_1965_69_3480, Viscoelastic_1971_9_209, Temperature_1980_12_43, Temperature_1982_20_729, Viscoelastic_1995_28_6432} We take this as a great relief since it is entirely unclear how such generalized models might be constructed.

Finally, we note that in addition to segmental fragility, there are many properties of polymer glass-forming liquids which have been claimed to be ``anomalous'' in comparison to non-polymeric glass-forming liquids. \cite{Connection_2016_659_133} For example, the ratio of $T_{\mathrm{g}}$ to the onset temperature $T_A$ for non-Arrhenius dynamics has been indicated to follow a different relationship between $T_{\mathrm{g}} / T_A$ and segmental fragility ($m_s$) than found in non-polymeric liquids, which is perhaps no surprise given the vastly different values of fragility often observed in polymer liquids. It is completely straightforward to calculate this temperature ratio, and, indeed, any ratio of characteristic temperatures that one might be interested in using the GET, and to compare these values with $m_s$, the extent of cooperative motion, or thermodynamic parameters of relevance for characterizing molecular packing frustration to better understand observed trends as considered in the present paper. There is a lot of interest in the thermodynamic scaling exponent, a basic measure of the anharmonicity of intermolecular interactions in liquids \cite{Equation_2021_54_3247} leading to the phenomenon of thermodynamic scaling in which the structural relaxation time $\tau_{\alpha}$, and many other dynamic properties, can be expressed in terms of a ``universal'' reduced variable, $TV^{\gamma}$, where $V$ is the material volume and $\gamma$ is a scaling exponent describing how $T$ and $V$ are linked to each other when either quantity is varied. This is another currently poorly understood property of interest that can be calculated systematically within the GET framework \cite{Thermodynamic_2013_138_234501} for essentially any polymer fluid to understand how this quantity is related to fragility and the extent of collective motion or thermodynamic properties that likewise strongly reflect anharmonic intermolecular interactions, such as the thermal expansion coefficient and isothermal compressibility. Despite the approximate nature of the GET, this model has proven to be a powerful tool for unraveling complex relations between molecular and thermodynamic variables on dynamic and thermodynamic properties of polymer materials.

\section*{Disclaimer}

Certain commercial equipment, instruments, software, or materials are identified in this paper to foster understanding. Such identification does not imply recommendation or endorsement by the National Institute of Standards and Technology, nor does it imply that the materials or equipment identified are necessarily the best available for the purpose.

%% classes. This should be given within the "acknowledgement"
\begin{acknowledgement}
W.-S.X. acknowledges the support from the National Natural Science Foundation of China (Nos. 22222307 and 21973089). X.X. acknowledges the support from the National Natural Science Foundation of China (No. 22373100) and the Youth Innovation Promotion Association of Chinese Academy of Sciences (No. 2023233). This research used resources of the Network and Computing Center at Changchun Institute of Applied Chemistry, Chinese Academy of Sciences.
\end{acknowledgement}

%%%%%%%%%%%%%%%%%%%%%%%%%%%%%%%%%%%%%%4%%%%%%%%%%%%%%%%%%%%%%%%%%%%%%%
\begin{suppinfo}

Discussion of the segmental fragility and extent of cooperative motion for polymer melts having variable cohesive interaction strength and applied pressure and the high temperature vibrational prefactor as a function of molecular and thermodynamic parameters.

\end{suppinfo}

\bibliography{refs}

\end{document}